\documentclass[pdflatex,sn-mathphys-num]{sn-jnl}

\usepackage{graphicx}%
\usepackage{multirow}%
\usepackage{amsmath,amssymb,amsfonts}%
\usepackage{amsthm}%
\usepackage{mathrsfs}%
\usepackage[title]{appendix}%
\usepackage{xcolor}%
\usepackage{textcomp}%
\usepackage{manyfoot}%
\usepackage{booktabs}%
\usepackage{algorithm}%
\usepackage{algorithmicx}%
\usepackage{algpseudocode}%
\usepackage{listings}%
\usepackage{mdframed}
\usepackage{booktabs}
\usepackage{tabularx}
\usepackage{array}
\usepackage{adjustbox} 
\usepackage{tcolorbox}
\newtcolorbox{rqbox}{
  colback=gray!10,
  colframe=gray!50,
  boxrule=0.5pt,
  arc=2pt,
  left=4pt,
  right=4pt,
  top=2pt,
  bottom=2pt,
  fonttitle=\bfseries,
}

\theoremstyle{thmstyleone}%
\theoremstyle{thmstyletwo}%

\theoremstyle{thmstylethree}%

\begin{document}

\title[Article Title]{Who Pays the Review Cost? Triage, Fairness, and Accountability in AI-authored Pull Requests}


\author*{\fnm{Md Shamimur} \sur{Rahman}}\email{mdr614@usask.ca}

\author{\fnm{Khairul} \sur{Alam}}\email{kha060@usask.ca}

\author{\fnm{Banani} \sur{Roy}}\email{banani.roy@usask.ca}
\author{\fnm{Chanchal K.} \sur{Roy}}\email{croy@cs.usask.ca}

\affil{\orgdiv{Department of Computer Science}, \orgname{University of Saskatchewan}, \city{Saskatoon}, \state{Saskatchewan}, \country{Canada}}




\abstract{AI coding agents are moving from local code assistance into pull-based workflows, where generated contributions must be reviewed, explained, and maintained within existing project norms. Although recent work has begun to characterize AI-authored pull requests (AIPRs), less is known about how reviewers govern their entry into review, how AI authorship reshapes credibility and fairness, and what intake mechanisms protect review sustainability. We report a mixed-method questionnaire survey of 239 practitioners from 31 countries with code-review experience and varying exposure to AIPRs. In the scenarios and self-reports elicited by the survey, AI authorship was not a categorical rejection signal. Instead, respondents described review effort as conditional on whether an AIPR arrived as an accountable contribution, with bounded scope, project-grounded rationale, validation beyond Continuous Integration (CI), contributor responsiveness, and identifiable post-merge ownership. This conditional logic extended to newcomer AIPRs, where respondents emphasized visible participation in the current review process over profile-level reputation alone. Qualitative responses further described shifts in mentoring, scrutiny, deferral, and routing when human stewardship was difficult to observe. We conceptualize missing rationale, validation, and ownership as \emph{reviewability debt}, the work reviewers must absorb when generated code lacks sufficient human grounding. These findings reframe AIPR governance around making human judgment observable before generated contributions consume scarce reviewer attention.
}




\maketitle
\section{Introduction}\label{sec:intro}

Modern Code Review (MCR) serves both as a quality gate and a socio-technical mechanism through which software projects allocate attention, transfer knowledge, negotiate responsibility, and decide whether a contribution deserves integration~\cite{bacchelli2013expectations,kononenko2016code,rigby2013convergent,beller2014modern}. In pull-based development, reviewers examine Pull Requests (PRs), interpret their intent, assess risks, request clarification, evaluate tests and supporting evidence, and decide whether the project can safely accept the change~\cite{gousios2014exploratory,tsay2014influence}. Prior work shows that review quality depends on technical factors such as review coverage, change size, tests, and reviewer expertise, but also on social factors such as prior contribution history, responsiveness, project familiarity, and author-reviewer interaction~\cite{mcintosh2016empirical,bosu2015characteristics,tsay2014influence,baysal2016investigating,sadowski2018modern}. This socio-technical role makes reviewer attention a limited project resource. Reviewers must decide not only whether a change is correct, but also whether the submitted contribution is reviewable enough to justify further investment. Large, unclear, poorly tested, or weakly explained PRs can shift effort from contributors to reviewers, increasing integration cost and delaying other work~\cite{mcintosh2016empirical,ebert2021exploratory,yu2015wait}. These costs are especially important in pull-based ecosystems, where contributors vary in familiarity, responsiveness, and long-term responsibility for the code they submit~\cite{gousios2014exploratory,tsay2014influence,steinmacher2015social,ruangwan2019impact}. Thus, code review depends on an implicit social contract: contributors make changes reviewable, and reviewers invest scarce attention in return.

AI-assisted coding changes this contract. Coding tools and agents such as Copilot Cloud \cite{github2026copilotCloudAgent}, Claude Code \cite{anthropic2026claudeCodeGithubActions}, Codex Web \cite{openai2026codexWeb}, and related systems can generate code, revise implementations, create branches, and prepare PRs with different levels of human involvement~\cite{bird2022taking,vaithilingam2022expectation,barke2023grounded,watanabe2025use,li2025rise,ehsani2026ai}. These tools can reduce the cost of producing a patch, but they do not necessarily reduce the cost of reviewing that patch~\cite{vaithilingam2022expectation,barke2023grounded}. A PR may be syntactically plausible, accompanied by a polished description, and supported by generated tests, while still leaving reviewers uncertain about project context, edge cases, maintainability, and post-merge ownership~\cite{barke2023grounded,perry2023users}. This creates a structural tension for code review: code production becomes cheaper, but review attention remains scarce \cite{kononenko2016code}. The key question, therefore, is who bears the review cost required to establish that generated code is acceptable for integration.

This paper studies that tension as a problem of \emph{reviewability and governance}. We define \emph{reviewability} as the extent to which a PR provides enough context, evidence, scope control, and accountable human engagement for reviewers to invest effort productively. Reviewability is different from correctness. A correct change may still be hard to review if it lacks rationale, test evidence, decomposition, or an accountable maintainer. Conversely, an incomplete change may be reviewable if it clearly explains intent, bounds risk, surfaces uncertainty, and supports constructive feedback. AI-authored PRs (AIPRs) make this distinction important because reviewers may no longer be able to infer human effort, understanding, or ownership from the submitted code alone~\cite{bird2022taking,barke2023grounded,li2025rise,ehsani2026ai}. Existing research provides useful foundations, but leaves this governance problem underexplored. MCR research has studied PR acceptance, review quality, reviewer expertise, confusion, reviewer recommendation, and contributor signals~\cite{bacchelli2013expectations,mcintosh2016empirical,kononenko2016code,ebert2021exploratory,thongtanunam2016revisiting}. Similarly, AI-assisted development research has examined how developers use generated code, how they validate suggestions, and how AIPRs appear in repositories~\cite{vaithilingam2022expectation,barke2023grounded,bird2022taking,watanabe2025use,li2025rise,ehsani2026ai}. However, less is known about how reviewers decide whether AIPRs deserve review effort, how low-effort AI-assisted submissions affect fairness and newcomer treatment, and what team-level rules reviewers expect when generated contributions enter normal review queues. 

To address this gap, we conducted an online survey and analyzed 239 valid responses from software practitioners with code review experience and exposure to AI-authored or agent-generated PRs. We examine AIPR review as a social and governance problem, focusing on how AI authorship changes review work, how reviewers decide whether generated contributions deserve attention, how fairness and newcomer credibility are negotiated, and what safeguards reviewers expect projects to adopt. Our results show that reviewers do not reject AIPRs categorically. Instead, they apply a conditional review model in which AIPRs become reviewable when they provide clear intent, bounded scope, concrete validation, responsive human engagement, and identifiable ownership. When these signals are missing, reviewers are more likely to defer, route, audit, or reject the contribution because the PR shifts explanation, validation, and ownership costs onto maintainers. These findings position AIPR governance as a workflow problem. Projects need mechanisms that make AI use, evidence, ownership, risk, and reviewer-bandwidth costs visible before generated contributions consume scarce human attention. Overall, this paper makes the following three main contributions:


\setlength{\leftmargini}{0.9em}
\begin{itemize}
    \item \textbf{A reviewability-centered account of AIPR triage.} We show that reviewers condition their willingness to proceed on evidence, scope, ownership, responsiveness, and risk-specific context, not on AI authorship alone.
    \item \textbf{An empirical characterization of the social contract of AIPR review.} We identify how visible human effort, newcomer behavior, fairness triggers, and effort asymmetry shape whether reviewers provide mentoring, constructive feedback, audit-like scrutiny, or rejection.
    \item \textbf{Governance implications for AI-assisted review workflows.} We synthesize respondent-proposed safeguards into workflow mechanisms for making AIPRs more reviewable and accountable, including AI-use boundaries, risk-specific evidence, queue controls, owner routing, and generated-diff containment.
\end{itemize}

\section{Background and Related Work}

\subsection{Signals and Decision-making in PR Review}

MCR has been widely studied as a lightweight yet consequential form of software quality assurance. Prior work shows that review coverage, reviewer participation, and reviewer expertise are associated with post-release quality and defect risk~\cite{mcintosh2016empirical}. Review quality also depends on feedback usefulness, reviewer familiarity, change quality, actionable comments, and information about rationale, expected behavior, dependencies, tests, and surrounding context~\cite{kononenko2016code,bosu2015characteristics,bacchelli2013expectations,pascarella2018information,spadini2018testing}. When such information is missing or ambiguous, reviewers can become confused and review progress can slow~\cite{ebert2021exploratory}.

Pull-based development adds contributor-context judgment because reviewers often evaluate changes from contributors who vary in project membership, prior interaction, and familiarity with project norms. As a result, PR acceptance and latency are influenced by change size, test status, discussion activity, contribution history, social connections, and other relationship-based signals~\cite{gousios2014exploratory,tsay2014influence,tsay2014let,gousios2015work,yu2015wait,baysal2016investigating}. Reviewer recommendation and ownership research further shows that expertise, subsystem familiarity, and prior review activity affect who is considered suitable to evaluate a change~\cite{thongtanunam2016revisiting,ruangwan2019impact}. Newcomer research similarly shows that contribution success depends on understanding project norms, responding to feedback, and building trust through interaction~\cite{steinmacher2015social,gousios2014exploratory,tsay2014influence}. Together, these works show that PR review relies on technical, social, and historical signals. However, these signals were studied mainly when submitted contributions were assumed to be authored and owned by humans. AIPRs complicate that assumption because contributor history, explanation quality, responsiveness, and ownership may need to be interpreted differently when substantial parts of a PR are perceived as generated. Our study builds on MCR research by examining how reviewers use these signals to decide whether AIPRs should receive review effort, be deferred, or be governed through additional safeguards.

\subsection{Newcomers, Casual Contributors, and Fairness in Review}

The social signals discussed above become particularly important for contributors who have little or no prior relationship with a project. Casual contributors constitute a substantial part of the contributor population even though their individual contribution volume is small. \citet{pinto2016more} found that casual contributors represented 48.98\% of contributors to popular GitHub projects but accounted for only 1.73\% of commits, with contributions ranging from documentation and typo fixes to bug fixes and new features. Not all attempted contributions become accepted contributions. \citet{steinmacher2018almost} studied ``quasi-contributors'' whose PRs were not accepted and identified duplicated or superseded contributions and mismatches with project direction among common reasons for nonacceptance. Approximately 30\% of surveyed quasi-contributors reported that the experience demotivated them or discouraged further contribution. PR abandonment can similarly result from both sides of the interaction: \citet{li2021you} found that lack of integrator response as well as contributors' loss of time or interest can cause PRs to stall or be abandoned, with consequences for project coordination and maintenance.

For contributors without an established project history, reviewers must often form judgments from the information that is observable in the current interaction. \citet{marlow2013impression} showed that developers form impressions of others on GitHub from activity traces and profile information. Such contributor-associated cues can also be associated with differences in PR outcomes. For example, \citet{terrell2017gender} found different acceptance patterns depending on contributor status and whether gender was identifiable, illustrating that evaluation in pull-based development can be associated with social information beyond the submitted code itself. These observations also raise questions about perceived fairness in review. Drawing on organizational justice theory~\cite{colquitt2001dimensionality}, \citet{german2018my} proposed a framework for studying fairness in MCR that distinguishes distributive, procedural, interpersonal, and informational dimensions. These dimensions concern review outcomes, the allocation of review opportunities, the consistency of review procedures, contributors’ experiences during review, and the transparency of review decisions.

AI-assisted contribution introduces an additional source of uncertainty into these judgments. A newcomer may use AI extensively while still understanding, validating, and taking responsibility for the submitted change, whereas another contribution may transfer much of the explanation and validation burden to maintainers. Consequently, perceived AI authorship alone may provide limited information about contributor commitment or reviewability. This motivates examining which observable signals reviewers use when deciding whether to invest effort in newcomer AIPRs, and under what circumstances authorship-dependent review practices are perceived as appropriate or unfair.

\subsection{AI-assisted Development and Agent-authored PRs}

AI-assisted programming research shows that developers use generated suggestions to accelerate familiar tasks, explore implementation options, and reduce routine coding effort~\cite{peng2023impact,vaithilingam2022expectation,barke2023grounded,bird2022taking}. Generated code, however, requires inspection, adaptation, debugging, and contextualization before safe use~\cite{vaithilingam2022expectation,barke2023grounded,bird2022taking}. It can also be sensitive to prompt variation, contain insecure patterns, or fail on tasks requiring project-specific reasoning and broader execution context~\cite{mastropaolo2023robustness,pearce2025asleep,fu2025security,corso2024generating}. Recent work has moved from local suggestions to agentic workflows, where coding agents can inspect repositories, modify files, create branches, and open or update PRs \cite{liu2025codexgraph, huang2023agentcoder, yang2024swe}. Empirical studies of AI-assisted and agent-authored PRs report variation in task type, change size, Continuous Integration (CI) outcomes, review activity, and merge decisions~\cite{watanabe2025use,li2026aidev,ogenrwot2026ai,ehsani2026ai,peralta2026agentic,watanabe2026ai}. These studies show that generated contributions are entering repository workflows, but provide less evidence on how reviewers decide whether such PRs are reviewable, whether the contributor supplied enough human validation, and what safeguards are needed before generated work consumes review effort. This paper uses \emph{Human-authored PRs} (HPRs) for PRs where a human is perceived as the primary author and accountable decision-maker, even with minimal AI support. We use \emph{AI-authored PRs} (AIPRs) for PRs perceived as substantially AI-produced, including agent-generated PRs submitted with varying levels of human supervision. This distinction matters because reviewers rarely know how each line was produced, so they infer authorship and accountability from observable cues such as disclosure, PR descriptions, tests, commit history, contributor responses, and signs of manual adaptation~\cite{pascarella2018information,liu2019automatic,tang2024study,tufano2026developers}. This perspective reflects ordinary review practice: reviewers proceed from visible artifacts, not complete knowledge of the human-AI production process.

\subsection{Human Review of Agent-authored PRs}
\label{sec}

While the preceding work establishes that agent-authored contributions are increasingly present in repository workflows, a smaller but rapidly growing body of research focuses specifically on how those contributions are reviewed by humans. Repository studies suggest that the review process surrounding AIPRs differs in both participation and interaction. \citet{duma2026these} found that many AI-generated PRs receive no review and that, among reviewed AIPRs, automated agents account for a substantial portion of review activity, while human involvement often takes the form of steering or supervising the agent rather than independently reviewing the submitted code. Examining the content of review feedback, \citet{haider2026understanding} analyzed 19,450 inline comments from 3,177 agent-authored PRs and identified 12 recurring themes. Beyond functional correctness, reviewers frequently raised concerns involving documentation, refactoring, code style, testing, and security. At a broader interaction level, \citet{zhong2026human} analyzed 278,790 review conversations and found that human reviewers engaged in 11.8\% more review rounds when reviewing AI-generated code than human-written code. How agents communicate through PR artifacts can also affect subsequent review interaction. \citet{watanabe2026ai} found systematic differences in how coding agents construct PR descriptions and reported associations between description characteristics and human review engagement and merge outcomes. Together, these studies indicate that review of agent-authored contributions cannot be understood solely from whether the resulting PR is eventually merged: the form of the contribution, the information accompanying it, and the resulting human--agent interaction also matter.

Automation can additionally operate on the reviewer side of the interaction. \citet{tufano2021towards} investigated automatically translating reviewers' natural-language comments into code changes and reported successful implementation for up to 31\% of comments in their evaluation. In an industrial setting, \citet{frommgen2024resolving} reported that developers at Google used machine-learning-generated edits to address 7.5\% of reviewer comments. Such systems demonstrate that parts of the review-response cycle can be automated, but they provide less evidence about the conditions under which reviewers themselves consider automated remediation appropriate, particularly when the original contribution is also AI-authored.

Repository traces provide important behavioral evidence about AIPR outcomes and observable review interactions, and qualitative inspection of those traces can recover some explicit decision rationales. However, they provide more limited access to reviewers' unexpressed criteria before investing effort, the circumstances under which they would defer or deprioritize a contribution, or their perceptions of responsibility, reciprocity, and fairness. Our study complements this repository-level evidence by directly eliciting reviewers' stated reasoning about these decisions.

\subsection{Human Oversight and Accountability in AI-mediated Work}
Research on trust in automation shows that automation can be misused, disused, or over-trusted when human reliance is not calibrated to system capability, task context, and available evidence~\cite{parasuraman1997humans,lee2004trust,hoff2015trust}. Human-AI interaction research similarly argues that AI systems should surface uncertainty, limitations, and control boundaries so users can decide when to rely on, question, or override~\cite{amershi2019guidelines,cai2019human,chen2023understanding}. For software development, generated code should be evaluated based on its technical correctness and on whether the development workflow supports human understanding, validation, and accountability. Emerging work on AI-assisted and agentic software development makes this oversight problem explicit by emphasizing the effects of agentic workflows on documentation, maintainability, security, coordination, organizational risk, and task completion~\cite{roychoudhury2025ai,dorner2025quo,zhong2026human,haider2026understanding}. However, existing work provides limited empirical evidence on how reviewers translate these concerns into concrete review decisions, credibility judgments, and team-level rules for AIPRs.

\subsection{Review Burden and Governance of AI-assisted Contributions}
\label{sec}

Recent studies have also examined how AI-generated contributions affect the reviewing capacity of software communities. Practitioner discussions increasingly describe an asymmetry between the cost of generating a contribution and the cost of evaluating it. \citet{baltes2026endless} analyzed 1,154 Reddit and Hacker News posts discussing low-quality AI-generated content, often described by practitioners as ``AI slop,'' and identified review friction as one of three major thematic clusters. Their analysis characterizes situations in which productivity gains for contributors can externalize validation and quality-assurance costs to reviewers and maintainers. \citet{baltes2026ai} develops this argument through a commons perspective, describing low-cost generation and comparatively expensive human review as a governance problem and discussing responses grounded in principles for managing shared resources.

Evidence of similar pressures appears at the repository and community level. \citet{afroz2026ai} combined practitioner accounts, longitudinal analysis of 294 repositories, interviews, and a survey of 229 Open Source Software (OSS) practitioners to study the effects of AI-generated contributions on open-source sustainability. Their analysis reports increasing contribution pressure alongside changes in merge outcomes, including an 18.18\% decrease in the merge rate of one-time contributors relative to their estimated counterfactual. The study also documents defensive practices that communities use to protect limited review capacity, while noting potential tensions between such controls and the openness traditionally associated with OSS contribution.

Projects are increasingly formalizing these concerns through AI contribution policies. \citet{hora2026ai} identified AI-related policies in 118 of the 1,000 most-starred GitHub repositories they examined. Among those policies, 78\% permitted AI-assisted contributions, 51\% required disclosure, and 74\% required human involvement. A subsequent analysis of 281 policies by \citet{hora2026we} similarly found that most projects permit some AI use while frequently imposing conditions involving human participation, accountability, and disclosure. \citet{robles2026you} examined the rationales behind such policies and identified concerns involving scarce maintainer attention, contributor accountability, meaningful review interaction, quality and legal risk, and preservation of pathways through which newcomers can learn and participate. At the repository level, \citet{chen2026making} found that adoption of explicit AI policies was associated with greater AI-use disclosure, richer review interaction, and differences in code-quality outcomes.

These aggregate patterns are also visible in individual project policies. Some projects restrict or reject AI-generated contributions because of provenance, quality, or accountability concerns~\cite{gentoo2024aipolicy,netbsd2024commit,qemu2025provenance}, whereas others permit AI assistance while making the human contributor responsible for understanding, validating, disclosing, and maintaining the resulting work~\cite{llvm2026aipolicy,fedora2025aipolicy}. Despite variation in implementation, these policies increasingly treat human stewardship and reviewer attention as resources that AI-assisted contribution workflows must explicitly protect. This work documents review burden at the community level and identifies the formal rules projects are adopting in response. It provides less evidence, however, about how individual reviewers operationalize such concerns in everyday decisions: which conditions must be satisfied before they invest review effort, how they adapt communication or checking practices, and when they perceive stricter handling of AIPRs as appropriate. Our study examines these reviewer-level judgments directly.

\subsection{Interpretive Theoretical Lenses}
\label{sec:lens}

We use three established theories retrospectively to interpret patterns in the survey results. These theories did not guide questionnaire construction, coding, or confirmatory hypotheses. Instead, they provide complementary explanations for the credibility, reciprocity, and fairness mechanisms observed in our findings. We return to these theoretical interpretations in Section~\ref{sec:discussion}.

\textbf{Signaling theory.} Signaling theory explains how observable cues communicate qualities that are otherwise difficult to assess~\cite{spence1978job,connelly2011signaling}. Software engineering research has similarly used signaling theory to explain how repository artifacts communicate otherwise unobservable project qualities \cite{trockman2018adding}. In AIPR review, descriptions, testing evidence, issue participation, responsiveness, and signs of manual adaptation can serve as signals of preparation and human engagement. Because generative AI can make some traditional signals easier to produce, this lens helps interpret why reviewers
may place greater weight on signals tied to observable human participation.

\textbf{Social exchange theory.}
Social exchange theory views social interaction as a reciprocal exchange of effort and resources~\cite{blau2017exchange}. Code review similarly involves an exchange in which contributors invest effort in preparing and explaining changes, while reviewers invest attention, feedback, and mentoring. This lens helps interpret the effort asymmetry, particularly how visible contributor effort relates to reviewers' willingness to reciprocate with deeper review or mentoring.

\textbf{Organizational justice theory.}
Organizational justice distinguishes distributive, procedural, interpersonal, and informational dimensions of perceived fairness \cite{colquitt2001dimensionality}. Prior work has applied this framework to MCR~\cite{german2018my}. We use it to interpret whether authorship-dependent review practices are perceived as fair when reviewer attention is allocated differently, procedures remain consistent, contributors are treated respectfully, and decisions are adequately explained.

\subsection{Positioning of This Study}

Prior work provides complementary evidence about AIPRs from repository outcomes, observed review interactions, maintainer burden, and project-level governance. Our study focuses on a different layer of the review process: reviewers' stated reasoning about when an AIPR deserves review effort, which signals make a generated contribution reviewable, and how triage decisions interact with fairness, reciprocity, and accountability. This perspective complements repository-based evidence by examining considerations that may shape review decisions before they become visible in repository traces. Table~\ref{tab:positioning} summarizes this distinction across the principal streams of related work.

\begin{table*}[t]
\centering
\caption{Positioning of this study relative to the principal streams of prior
work. ``Central'' indicates a primary analytical focus, ``Partial'' indicates
that the aspect is addressed but is not central, and ``Limited'' indicates
that it receives little direct attention.}
\label{tab:positioning}
\footnotesize
\renewcommand{\arraystretch}{1.15}

\begin{tabularx}{\textwidth}{
@{}
>{\raggedright\arraybackslash}p{0.20\textwidth}
>{\raggedright\arraybackslash}p{0.20\textwidth}
>{\raggedright\arraybackslash}p{0.13\textwidth}
>{\raggedright\arraybackslash}p{0.13\textwidth}
>{\raggedright\arraybackslash}p{0.13\textwidth}
>{\raggedright\arraybackslash}X
@{}
}
\toprule
\textbf{Research stream} & \textbf{Primary evidence} & \textbf{Reviewer reasoning} &
\textbf{Pre-review triage} & \textbf{Fairness / reciprocity} & \textbf{Governance / accountability} \\
\midrule

Traditional PR and newcomer studies
~\cite{gousios2014exploratory,tsay2014influence,
steinmacher2018almost,german2018my} & Repository data, surveys, interviews & Partial
& Partial & Partial--Central & Partial
\\

Agent-authored PR outcome studies
~\cite{watanabe2025use,ehsani2026ai,
peralta2026agentic,ogenrwot2026ai} & Repository traces & Partial & Limited & Limited
& Partial
\\

Human review of agent-authored PRs
~\cite{duma2026these,haider2026understanding,
zhong2026human,watanabe2026ai} & Review traces and conversations & Partial & Limited
& Limited & Partial
\\

Maintainer-burden studies
~\cite{baltes2026endless,baltes2026ai,
afroz2026ai} & Public discussions, repositories, interviews, surveys & Partial &
Partial & Partial & Central
\\

AI contribution-policy studies
~\cite{hora2026ai,hora2026we,
robles2026you,chen2026making} & Project policies and repository data & Limited
& Partial & Partial & Central
\\

\midrule

\textbf{This study} & \textbf{Survey of 239 practitioners} & \textbf{Central} & \textbf{Central} & \textbf{Central} & \textbf{Central}
\\

\bottomrule
\end{tabularx}
\end{table*}

\section{Methodology}

\subsection{Research Questions}

We organize the study around how AIPRs reshape review work, reviewability triage, and governance expectations. Our study is guided by the following Research Questions (RQs):

\begin{itemize}
    \item \textbf{RQ$_1$: How do AIPRs change the review work that reviewers must manage?} We examine whether AIPRs alter communication expectations, review-volume pressure, low-effort submission patterns, and review practices adopted or stopped after AIPRs became common.
    
    \item \textbf{RQ$_2$: What makes an AIPR reviewable enough to receive reviewer effort?} We study how reviewers triage AIPRs before detailed review, including willingness to proceed when CI passes, stricter entry criteria, deferral triggers, suspected AI-assisted drive-by PRs, and newcomer credibility signals.
    
    \item \textbf{RQ$_3$: How do reviewers expect AIPR workflows to preserve fairness and accountability?} This RQ focuses on the social and governance norms reviewers use to make AIPR review sustainable. We analyze reviewer-proposed team guidelines, effort-asymmetry concerns, and fairness triggers that shape whether reviewers provide mentoring, constructive feedback, audit-like scrutiny, or rejection.
\end{itemize}

\subsection{Survey Design and Data Analysis Plan}
\label{sec:survey-method}

We addressed the RQs through a cross-sectional mixed-method questionnaire survey. The survey design, recruitment, and analysis followed established guidance for empirical survey research in software engineering~\cite{pfleeger2001principles,kitchenham2008personal} and
the ACM SIGSOFT Empirical Standards~\cite{ralph2020empirical}. The questionnaire combined quantitative items for characterizing and comparing reviewer judgments with open-ended questions for examining the reasoning underlying those judgments.

\subsubsection{Target Population and Sampling Objective}

Our target population comprises software practitioners with code-review
experience in development settings where AI-authored or agent-generated pull requests are relevant. Because no comprehensive sampling frame of such practitioners exists, we used complementary recruitment sources to reach reviewers with different levels of professional experience, project context, and exposure to AIPRs. In particular, recruitment included developers identified from repositories with known agent-authored PR activity, developers from mature and active open-source projects, and practitioners reached through professional networks.

Our sampling objective was to obtain a heterogeneous practitioner sample capable of providing experience-based evidence about AIPR review rather than to reproduce the demographic distribution of the broader software-development population. Accordingly, the study examines patterns in reviewers' reported practices, judgments, and reasoning concerning reviewability, triage, fairness, and governance. Where HPR and AIPR items were paired, we additionally examine within-respondent differences while holding stable the respondent's broader review experience and disposition. We characterize the achieved sample and its coverage boundaries explicitly and restrict our conclusions to the contexts supported by the recruited participants~\cite{ralph2020empirical, baltes2022sampling}.

\subsubsection{Instrument Construction and Validation}
\label{sec:survey-design}

We developed the questionnaire iteratively following established guidance for empirical survey research in software engineering \cite{pfleeger2001principles,linaaker2015guidelines,molleri2016survey}. The initial questionnaire was constructed from our RQs and prior research on MCR, pull-request decision-making, newcomer participation, and AI-assisted software development. In particular, the candidate response alternatives for closed-ended and multiple-selection questions were informed by factors repeatedly identified in prior work, including change size and scope, testing and CI status, clarity of rationale and contextual information, contributor responsiveness, project familiarity, prior participation, and code ownership \cite{gousios2014exploratory,tsay2014influence,kononenko2016code,
bacchelli2013expectations,pascarella2018information,spadini2018testing,steinmacher2015social,ruangwan2019impact}. We adapted these established factors to the AIPR setting and introduced additional alternatives specific to AI-assisted contribution, such as visible human stewardship, manual adaptation of generated code, AI-generated boilerplate, and responsibility for validating and maintaining the resulting change.

We then assessed the questionnaire through two stages of pre-deployment validation. First, two industry practitioners with 8+ and 17+ years of professional experience and two software-engineering researchers with 11+ and 6+ years of MCR experience independently reviewed the questionnaire for relevance, clarity, terminology, and coverage of the intended review scenarios. We revised the wording, response alternatives, and structure iteratively based on their feedback. Second, we conducted a pilot study with 11 software practitioners with code-review experience. The pilot was used to assess whether the questions and response alternatives were understandable in realistic review contexts, whether important response possibilities were missing, and whether the questionnaire flow and length were appropriate. Feedback from the pilot was used to refine ambiguous wording, revise or clarify response alternatives, and finalize the questionnaire before deployment. Responses from the pilot study were not included in the reported dataset.

The resulting questionnaire is a literature-informed instrument developed and refined through iterative validation, without relying on a previously established psychometric scale. The predefined response alternatives capture plausible reviewer considerations identified through prior research and pre-deployment validation, but they are not intended as an exhaustive taxonomy. Most multiple-selection questions also included an \emph{Other} option so that respondents could report considerations not represented by the predefined alternatives. Accordingly, we interpret option-level results as respondents' endorsement of the alternatives presented in the questionnaire, while the open-ended responses provide a complementary source of evidence for
considerations outside the predefined response space.

\subsubsection{Questionnaire Structure and Measures}
\label{sec:questionnaire-structure}

The final questionnaire combined closed-ended, multiple-selection, ordinal-rating, paired-comparison, and open-ended questions. We used these formats for complementary purposes: structured items support systematic comparison across respondents and conditions, whereas open-ended questions allow respondents to explain their reasoning and introduce considerations not represented by the predefined alternatives. Table~\ref{tab:survey-structure} summarizes the questionnaire blocks analyzed in this paper, their response formats, and their relationship to the research questions. The complete questionnaire, including the original wording, response alternatives, and item order, is provided in the supplementary material.

\begin{table*}[t]
\centering
\caption{Structure of the questionnaire used in this study.}
\label{tab:survey-structure}
\small
\renewcommand{\arraystretch}{1.15}

\begin{tabularx}{\textwidth}{
@{}
>{\raggedright\arraybackslash}p{0.20\textwidth}
>{\raggedright\arraybackslash}p{0.19\textwidth}
>{\raggedright\arraybackslash}X
>{\raggedright\arraybackslash}p{0.09\textwidth}
@{}
}
\toprule
\textbf{Survey block} &
\textbf{Question type} &
\textbf{Purpose} &
\textbf{RQ} \\
\midrule

Participant background
&
Closed-ended and demographic
&
Characterize country, software-development experience, code-review
experience, and self-reported exposure to AI-authored or agent-generated PRs.
&
Context
\\\cmidrule{2-4}

Communication and review workload
&
Ordinal, categorical, and open-ended
&
Examine reported differences in review communication, changes in review
volume, and experiences with low-effort AI-generated submissions.
&
RQ$_1$
\\\cmidrule{2-4}

Willingness to proceed across change types
&
Paired ordinal ratings
&
Compare, within the same respondent, willingness to proceed with HPRs and
AIPRs without additional explanation when CI passes across eight change types.
&
RQ$_2$
\\\cmidrule{2-4}

Review-practice adaptation and team guidelines
&
Open-ended and multiple-selection
&
Identify practices respondents report adopting or discontinuing, desired
changes to team guidelines, and conditions for allowing an AI agent to propose
fixes for review comments.
&
RQ$_1$, RQ$_3$
\\\cmidrule{2-4}

Review entry and deferral
&
Multiple-selection
&
Examine the presented conditions respondents use when deciding whether to
begin, defer, or skip detailed review of an AIPR.
&
RQ$_2$
\\\cmidrule{2-4}

Newcomer credibility and suspected drive-by submissions
&
Multiple-selection and ordinal ratings
&
Examine reported handling of suspected AI-assisted drive-by PRs and the
influence assigned to predefined participation, validation, scope, and
engagement signals.
&
RQ$_2$, RQ$_3$
\\\cmidrule{2-4}

Fairness, reciprocity, and effort asymmetry
&
Open-ended
&
Elicit situations that heighten respondents' perceived obligation to treat a
submitter carefully or fairly and explore how perceived contribution-effort
differences relate to review effort and reciprocity.
&
RQ$_3$
\\

\bottomrule
\end{tabularx}
\end{table*}

The paired willingness block presented the same eight change types under HPR and AIPR conditions: low-risk cleanup, documentation updates, test additions for existing behavior, behavior-preserving refactoring, performance changes,
concurrency or asynchronous logic, authentication/security-sensitive code, and complex new-feature logic. In both conditions, respondents were asked to assume that CI passed. Holding the change type and CI assumption constant
supports a within-respondent comparison of whether perceived authorship is associated with different willingness to proceed without additional explanation.

The AIPR-specific blocks capture different stages of review. Entry-criteria items concern conditions respondents report applying before beginning detailed review, whereas deferral items concern characteristics that make them more
likely to postpone or skip review. The newcomer block separately examines how respondents report handling suspected AI-assisted drive-by submissions and how influential they consider predefined signals such as prior issue participation, response to CI failures, testing evidence, appropriate scope, manual adaptation of generated code, and professional response to feedback. Open-ended questions throughout the questionnaire provide complementary qualitative evidence about communication changes, review-practice adaptation, team guidelines, fairness, and perceived effort asymmetry.

\subsubsection{Recruitment and Sampling Frame}
\label{sec:recruitment}

We used a purposive, multi-frame recruitment strategy to reach practitioners with relevant experience in code review. First, we targeted reviewers from the AIDev dataset~\cite{li2026aidev}, which contains repositories with PRs created by coding agents including Claude, Codex, Devin, and Copilot. We retained projects with at least 500 stars and identified developers who had posted at least one review comment, yielding 516 publicly available email addresses. Second, we broadened the repository-based sampling frame beyond projects with known AIPR activity. We selected 133 active Apache Software Foundation projects and 49 additional mature GitHub projects using project activity, popularity, contributor count, commit history, and presence of active PRs as selection criteria. The selected projects covered several widely used programming languages, including Java, Python, C, C++, and JavaScript, and we verified that projects did not overlap across the repository sets. This broader repository-based process identified more than 4,000 eligible developers, of whom 3,357 had publicly available email addresses. Detailed project-selection criteria are reported in the supplementary material.

Finally, we disseminated the survey through LinkedIn to broaden participation beyond developers identified through public repository histories. Combining repository-derived recruitment with professional-network dissemination was intended to reduce dependence on a single sampling frame and to capture practitioners working across different development contexts. The survey was administered through SurveyMonkey\footnote{\url{https://www.surveymonkey.com/}}.

\subsubsection{Response Screening and Final Sample}
\label{sec:response-screening}

The survey received 261 response records during the two-month data-collection period. Before analysis, we screened the records for consent and questionnaire completion. We excluded one response for which consent was not provided and 21 substantially incomplete responses, resulting in a final dataset of 239 participants. We classified a response as substantially incomplete when more than 50\% of the questions were unanswered. Optional and open-ended questions were analyzed using the number of valid responses available for each item, and we report item-specific denominators when they differ from the full sample.

Since recruitment combined direct email invitations with open dissemination through LinkedIn, the total number of individuals who encountered the survey is not observable. We therefore report the observed response disposition instead of a single invitation-level response rate. Participant characteristics and item-level denominators are reported in the Results and supplementary material.

\subsubsection{Quantitative Analysis}
\label{sec:analysis-plan}

We matched the quantitative analysis to the measurement scale and dependency structure of each survey item. Descriptive statistics are the primary evidence for single-condition questions: for categorical and multiple-selection items, we report valid counts and percentages, and for ordinal items we report response distributions and summary statistics appropriate to ordered data.

For questions that directly compared HPR and AIPR conditions within the same respondent, we treated observations as paired. In particular, willingness to proceed across change types was analyzed using Wilcoxon signed-rank tests, with matched-pairs rank-biserial correlation reported as an effect-size measure \cite{wilcoxon1992individual,kerby2014simple}. We additionally report the proportion of respondents whose AIPR rating was lower than, equal to, or higher than the corresponding HPR rating to make the direction of within-participant change transparent. For multiple-selection questions, each response alternative was represented as a binary indicator and selection rates were reported independently. We treat these rates as endorsement of the response alternatives presented in the questionnaire rather than as an exhaustive ranking of all possible reviewer considerations. For the newcomer-credibility matrix, in which each participant rated the same set of signals on a common ordinal scale, we used the Friedman test
\cite{friedman1937use} with Kendall's $W$ as an effect-size measure \cite{kendall1939problem} to assess overall within-respondent differences among
the presented signals. Where post-hoc paired comparisons were examined, we used Wilcoxon signed-rank tests.

Exploratory analyses examined whether selected patterns varied with respondents' AIPR exposure, software-development experience, or code-review experience. Depending on the outcome type, we used Spearman's rank correlation \cite{spearman1961proof}, Kruskal--Wallis tests \cite{kruskal1952use}, Fisher's exact tests \cite{fisher1922interpretation}, or logistic regression \cite{hosmer2013applied}. These analyses are treated as contextual rather than confirmatory. In addition, some analyses involved multiple related comparisons addressing the same construct, such as the paired HPR--AIPR comparisons across change types and the post-hoc comparisons among newcomer-credibility signals. Testing each comparison independently at $\alpha=.05$ would increase the probability of at least one false-positive result within such a family. We therefore controlled the family-wise error rate within each set of related comparisons using the Holm procedure~\cite{holm1979simple}. Holm's sequential procedure provides strong control of the family-wise error rate while being less conservative than the standard Bonferroni correction. All tests were two-sided with $\alpha=.05$. We interpret statistical significance together with effect sizes and response distributions rather than relying on $p$-values alone.

\subsubsection{Qualitative Analysis}
\label{Analysis Plan}

We analyzed the open-ended responses using codebook-based thematic coding \cite{braun2006using}. We use this term to distinguish our procedure from reflexive thematic analysis, because our analysis employed an explicit codebook, independent coders, and inter-rater agreement.

Two researchers with 10+ and 8+ years of software-development experience first independently examined a random sample of 100 responses. They identified recurring concepts, compared the independently generated categories, and iteratively consolidated them into a shared codebook using a card-sorting procedure~\cite{zimmermann2016card}. Categories were defined with descriptions and decision rules, and responses could receive multiple codes when they expressed more than one concept. After stabilizing the codebook, the two coders independently applied it to the
remaining responses. For open-ended items with fewer than 20 valid responses, both coders examined all available responses. We measured inter-rater agreement on independently coded samples using Cohen's $\kappa$ \cite{cohen1960coefficient}. Disagreements were resolved through discussion and cases that remained unresolved were adjudicated by a third evaluator with more than 20 years of software-engineering experience.

The final codes were subsequently grouped into higher-level themes where
conceptually related categories described the same broader review phenomenon. Because multi-label coding was permitted, theme percentages are not mutually exclusive. We use the qualitative responses to explain, contextualize, and occasionally qualify the quantitative patterns rather than to estimate population prevalence.
\section{Evaluation and Results}

\subsection{Overview of Participant Demographics}

Over two months, the survey received 261 responses. After excluding 22 responses, one without consent and 21 substantially incomplete, we analyzed 239 valid responses. Participants came from 31 countries, with the largest groups from the United States (19.7\%, 47), Canada (15.5\%, 37), Bangladesh (10.9\%, 26), Germany (10.9\%, 26), and India (7.1\%, 17). The top five countries accounted for 64.0\% of responses, while the top 10 accounted for 83.7\%, indicating geographic breadth with some concentration in major respondent countries.

\begin{figure}[htbp]
\centering
\includegraphics[width=0.9\linewidth]{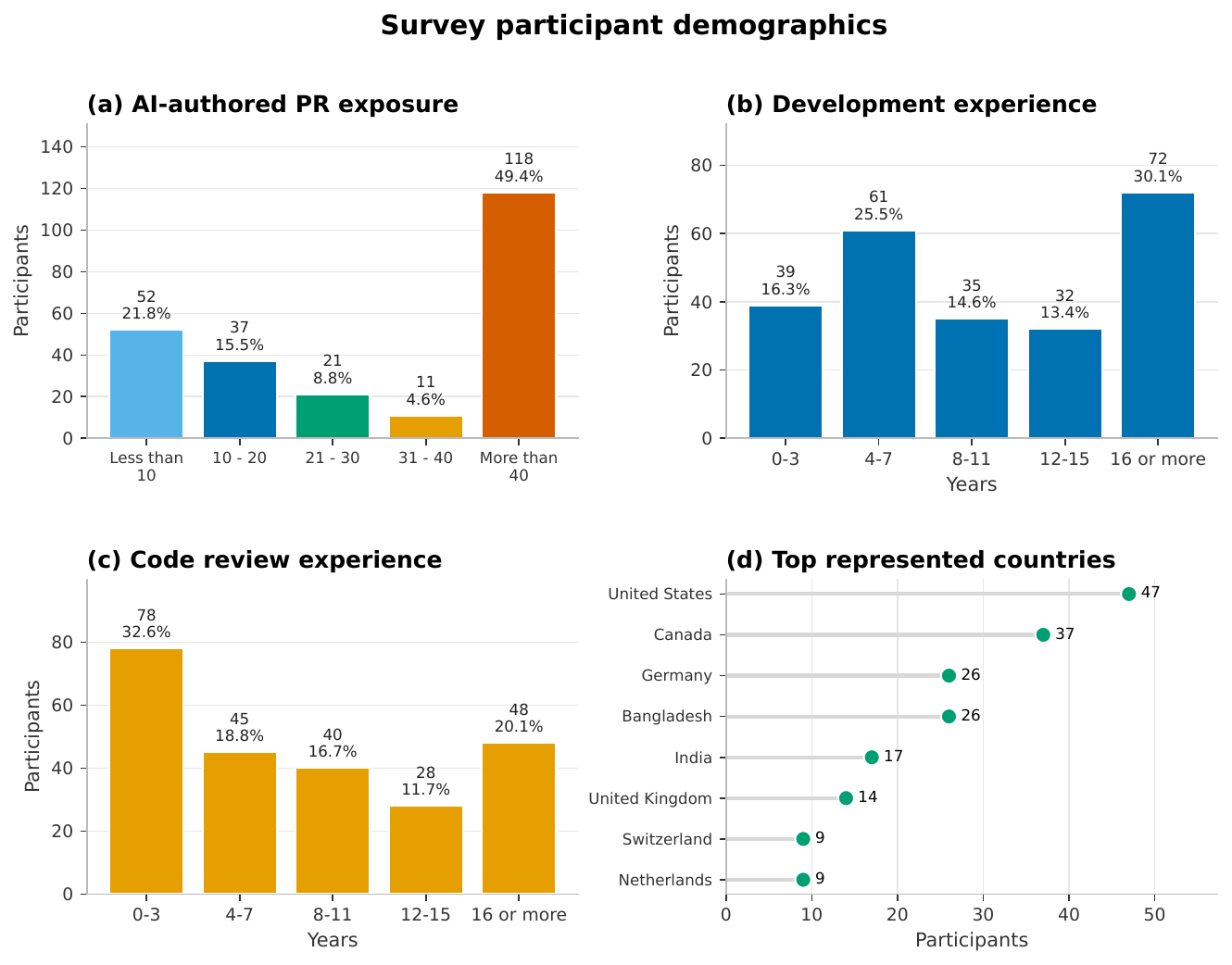}
\caption{Participant demographics. (a) shows participants' self-reported exposure to reviewing AI-authored or agent-generated PRs; (b) and (c) summarize software development and code review experience, respectively; (d) shows the eight most represented countries in the analyzed sample. Percentages are computed over the 239 valid responses.}
\label{fig:demographics}
\end{figure}

Participants varied in professional experience. Most had substantial software development experience, with 58.2\% (139) reporting more than seven years and 30.1\% (72) at least 16 years. Code review experience was also broad, with 48.5\% (116) reporting more than seven years and 32.6\% (78) reporting 0--3 years. The sample therefore includes many experienced developers and reviewers while retaining early-career perspectives. Participants also reported meaningful exposure to AIPRs: 78.2\% (187) had reviewed at least 10 AIPRs, 62.8\% (150) more than 20, and 49.4\% (118) more than 40. Overall, the sample is grounded in practical AIPR review experience while still including respondents with more limited exposure. Figure~\ref{fig:demographics} shows the full respondent distributions. 

\subsection{RQ$_1$- Review Work Under AI Authorship}
\label{sec:rq1}

RQ$_1$ examines how AIPRs reshape review work before approval decisions, including communication expectations, review-volume pressure, low-effort submission patterns, and practices reviewers adopted after AIPRs became common.

\subsubsection{Communication and workload conditions}

\begin{figure}[htbp]
\centering
\includegraphics[width=\linewidth]{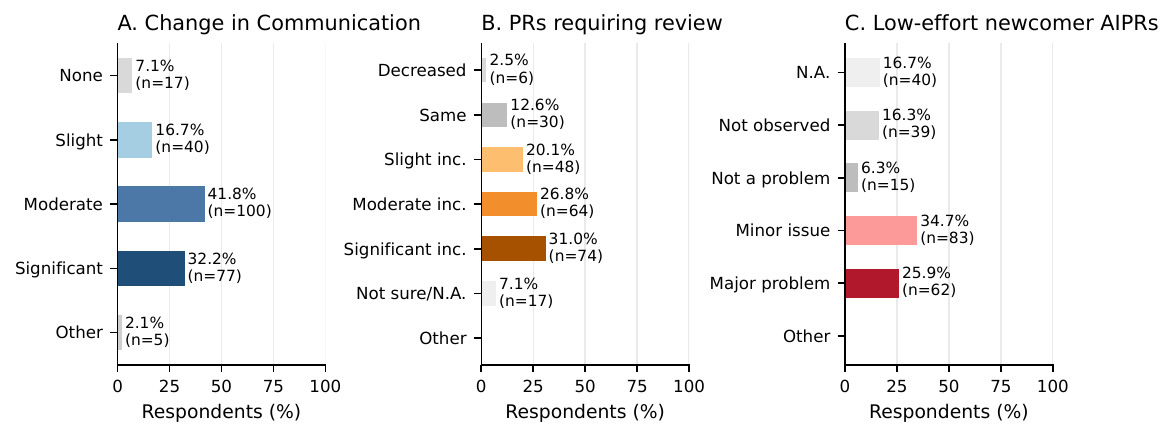}
\caption{Reviewers’ communication and workload adaptations to AIPRs.}
\label{fig:rq1-communication-workload}
\end{figure}

Figure~\ref{fig:rq1-communication-workload} shows that reviewers perceived AIPRs as changing both communication and workload. For communication expectations, 90.8\% reported at least some change and 74.1\% reported a moderate or significant change (Figure~\ref{fig:rq1-communication-workload}A). Responses were concentrated in the moderate and significant change categories. The 95\% confidence intervals were 86.5--93.8\% for any change and 68.2--79.2\% for moderate or significant change. This suggests that AIPRs are not reviewed under the same interaction norms as HPRs, since reviewers expect different forms of explanation, accountability, or engagement when a contribution is perceived as AI-authored.

\begin{table*}[htbp]
\centering
\caption{Themes in open-ended responses explaining how communication changes when reviewing AIPRs. Percentages are over 195 substantive responses and a response can carry several themes.}
\label{tab:rq1-communication-themes}
\resizebox{\linewidth}{!}{%
\begin{tabular}{p{0.20\linewidth}p{0.40\linewidth}p{0.60\linewidth}}
\toprule
\textbf{Theme} & \textbf{Theme context} & \textbf{Example responses} \\
\midrule

Validation-first communication (n = 117, 60.0\%) &
Responses describe asking for tests, edge cases, manual validation, logs, benchmarks, local runs, security checks, or other artifacts because generated code can look correct while missing important behavior. &
\textit{``[....] but difference starts when the PR looks pasted from a tool. Then I ask who reviewed the generated code and whether they tested it.''}, \textit{``[....] more direct about rejecting AI PRs that do not show human review. [....] I am reviewing the submitter's work, not the model's output, so they need to stand behind it.''}\\
\cmidrule{2-3}

Reconstructing missing intent and context (n = 74, 37.9\%) &
It captures AIPRs as lacking tacit project knowledge, business context, architectural history, or an intelligible rationale, making reviewers ask for intent and context before approval. &
\textit{``[......] in both positive and negative ways. In case of human submission, it is easy to infer the thought process, trade-off consideration, and level of understanding.''}, \textit{``AI does not have tacit knowledge, a human does. Forces me to front load a lot of tacit knowledge as context, over and over again.''} \\
\cmidrule{2-3}

Human ownership and explicit accountability (n = 56, 28.7\%) &
Responses report shifting attention from the model to the accountable human submitter, including whether the person understands, edited, disclosed, can explain, and will maintain the generated change. &
\textit{``AI generated code make me more explicit about accountability. I ask the person submitting it to confirm they understand the change and will maintain it.''}, \textit{``[...] ask the submitter to take responsibility for the generated code. want to know they check it, not just that the tool produced something that passes tests.''} \\
\cmidrule{2-3}

Tone shifts from mentoring to transactional review (n = 54, 27.7\%) &
It summarizes communication becoming more direct, checklist-like, or less mentoring-oriented for AIPRs, while human-authored reviews remain more collaborative and educational. &
\textit{``Human reviews require empathy and explaining things so the author learns. AI reviews are purely transactional. I can be totally blunt. [...]''}, \textit{``I usually communicate more directly on AI-authored PRs. With human-authored PRs, I explain the reasoning and try to teach the pattern. [...]''} \\
\cmidrule{2-3}

Conditional differences under ownership (n = 48, 24.6\%) &
Responses emphasize no change, similar standards, or positive and efficient communication when a trusted human owns the AI output and provides adequate context, tests, and rationale. &
\textit{``All PRs where I work are ultimately submitted by a human. Thus the human is responsible in the same way [....] whether it is generated with AI or not.''}, \textit{`` The difference is not always negative. Some AI-assisted PRs from experienced developers are very well prepared, and then communication is efficient [.....].} \\
\cmidrule{2-3}

Reviewability and governance concerns (n = 43, 22.1\%) &
This theme describes generated PRs as broad, verbose, noisy, low-effort, or costly to review, leading reviewers to ask for smaller scope, clearer boundaries, or to stop review until entry conditions are met. &
\textit{``Usually I communicate more explicitly. If an AI-generated PR is broad, vague, or unowned, I tell the submitter that it is not reviewable in its current form.''}, \textit{``AI-generated PRs in open source has been wholly negative. The amount of spam and verbosity of individual PRs has increased greatly. Projects I maintain have adopted a policy for this.''} \\

\bottomrule
\end{tabular}%
}
\end{table*}

Respondents also reported increased review-volume pressure. Since AIPRs became common in their projects, 77.8\% reported an increase in PRs requiring review, including 57.7\% who reported a moderate or significant increase (Figure~\ref{fig:rq1-communication-workload}B). The 95\% confidence intervals were 72.1--82.6\% and 51.4--63.8\%, respectively. Respondents also distinguished volume from workload quality: 60.7\% characterized low-effort AI-generated newcomer PRs as a minor issue or major problem, including 25.9\% who described them as a major problem (Figure~\ref{fig:rq1-communication-workload}C). The 95\% confidence intervals were 54.4--66.6\% and 20.8--31.8\%, respectively. Optional comments were few (5, 2.1\%), but they reinforced the same interpretation: communication changed most when AI-authored submissions appeared unaccountable, uncontextualized, or merely repeated tool output. Together, these results show that AIPRs reshape review practice by changing both the amount and kind of review work that maintainers must manage.

\begin{rqbox}
\textbf{Finding 1.1:} AIPRs changed review communication and workload conditions. Most respondents reported changed communication expectations, increased PR volume, and low-effort newcomer AIPRs as a review-pressure problem.
\end{rqbox}

\begin{table*}[t]
\centering
\caption{Themes in open-ended responses describing review practices adopted or stopped after AIPRs became common. Percentages are over 200 substantive responses and a response can carry several themes.}
\label{tab:rq1-practice-themes}
\resizebox{\linewidth}{!}{%
\begin{tabular}{p{0.20\linewidth}p{0.40\linewidth}p{0.60\linewidth}}
\toprule
\textbf{Theme} & \textbf{Theme context} & \textbf{Example responses} \\
\midrule

Active validation over surface trust (n = 125, 62.5\%) &
Respondents adopted deeper checks of tests, local execution, business logic, edge cases, security, and production behavior because AI-generated code can look clean while still being wrong. &
\textit{``I now routinely run AI changes locally and inspect the tests myself [....]. They often look right, pass tests and CI but fails on edge cases or references APIs that do not exist, [...].''}, \textit{``I stopped glancing over boilerplate and test additions and started reviewing them with maximum scrutiny. AI generates massive amounts of tests instantly, but it frequently hallucinates mock data, tests tautologies, or misses critical edge cases while making coverage look perfect.''} \\
\cmidrule{2-3}

Intent, provenance, and ownership gates (n = 119, 59.5\%) &
Respondents ask for human rationale, AI-use disclosure, generated-versus-manual provenance, PR-description consistency, and a clear human owner before review. &
\textit{``I now ask who will own the change after merge. If a PR is mostly AI-generated but no human can explain or maintain it, I do not treat it as review-ready.''}, \textit{``[..] ask what was AI-generated and what was manually edited. It helps me decide where to focus the review and whether the submitter actually shaped the final solution.''} \\
\cmidrule{2-3}

Reviewability and bandwidth controls(n = 90, 45.0\%) &
Respondents manage generated PR load by asking for smaller scope, issue-first submissions, templates, queueing, rejection, or pausing review until basic entry conditions are met. &
\textit{``I no longer review broad generated refactors in one pass. I ask for smaller staged PRs because the review cost grows very quickly when the change is mechanical but spread across many files.''}, \textit{``I started using a short template response for obvious low-effort AI PRs. It asks for tests, issue context, and confirmation that the contributor understands the change.''} \\
\cmidrule{2-3}

AI-assisted review triage (n = 18, 9.0\%) &
Some respondents adopted AI tools as pre-reviewers, secondary reviewers, triage aids, or consistency checkers so that human reviewers can focus on design, architecture, and risk. &
\textit{``Added code-review instructions and run an AI model before the human review to catch the most common issues. Then human review can focus on design, architecture, and big picture.''}, \textit{``Using AI to review the PR looking for consistency beyond the current repository.''} \\
\cmidrule{2-3}

Procedural feedback over mentoring (n = 12, 6.0\%) &
Respondents describe using shorter checklist-like responses, fewer educational comments, or more command-like feedback when a PR appears to be unmodified AI output. &
\textit{``I stopped writing long educational comments for PRs that look like unmodified AI output. I still help genuine newcomers, but I first ask them to show they understand the change.''}, \textit{``I started giving one clear checklist comment rather than many inline comments on low-effort AI submissions. It is fairer to the contributor and less exhausting for reviewers.''} \\

\bottomrule
\end{tabular}
}
\end{table*}

\subsubsection{Communication shifts toward evidence and accountability}

To explain how communication expectations changed, we thematically coded the 239 open-ended responses using the multi-label procedure described in Section~\ref{Analysis Plan}. The initial coding produced 12 groups, which were merged into six higher-level themes for reporting. Inter-coder agreement on the independently coded sample was substantial ($\kappa=0.89$). Overall, 195 responses (81.6\%) were assigned to at least one substantive theme, while 44 responses were too brief, unclear, neutral-only, or indicated no relevant AIPR review experience.

Table~\ref{tab:rq1-communication-themes} shows that AIPRs changed the communicative purpose of review. Reviewers shifted from conversational feedback grounded in author intent toward evidence-gated, accountability-focused interaction. The dominant theme was concrete validation, followed by concerns about missing intent, project context, and human ownership. Respondents reported minimal difference when a trusted human clearly owned the AI output. Still, many described AIPRs as requiring more explicit proof, narrower scope, and clearer accountability before review could proceed productively. As a secondary check, we tested whether communication orientation (positive, neutral/mixed, or negative) differed by respondents' code-review experience. The association between code-review experience and communication orientation was not significant ($\chi^2(8)=6.29$, $p=.615$, Monte Carlo $p=.623$, and Cramer's $V=0.12$). This suggests that evidence-gated communication was not driven mainly by junior or senior reviewers, but reflected a broadly shared response to AIPRs.

\begin{rqbox}
\textbf{Finding 1.2:} Communication around AIPRs shifted from conversational review toward evidence-gated accountability, with reviewers asking for concrete validation, reconstructed intent and context, explicit human ownership, and clearer reviewability conditions before engaging deeply.
\end{rqbox}

\subsubsection{Reviewers adopt concrete safeguards for AIPR work}

We next examined whether communication shifts translated into concrete review-practice changes. Respondents described practices they adopted or stopped following AIPR exposure. The initial coding of 239 responses produced 12 subthemes, which were merged into five higher-level themes. Inter-coder agreement was substantial ($\kappa=0.89$). Overall, 200 responses (83.7\%) were assigned to at least one substantive practice-adaptation theme, while 39 (16.3\%) indicated no substantive change, were unclear, or contained limited information.

Table~\ref{tab:rq1-practice-themes} shows that reviewers described concrete safeguards, not only changed attitudes. The most common adaptation was replacing surface-level trust with active validation, followed by intent, provenance, and ownership gates, and reviewability or bandwidth controls. These themes often co-occurred, suggesting a compound safeguard: establish technical evidence, confirm accountable human ownership, and then decide whether the AIPR warrants deeper review. Exploratory checks found no robust association between these themes and reported AIPR pressure (Figure~\ref{fig:rq1-communication-workload}), code-review experience, or AIPR exposure. Given the limited power of these checks, the findings only suggest that these safeguards were not confined to a specific pressure or experience group.

\begin{rqbox}
\textbf{Finding 1.3:} Reviewers responded to AIPR-related communication and workload pressure with safeguards. Most reported active validation, ownership and provenance gates, and reviewability or bandwidth controls, with adaptations broadly distributed across respondents.
\end{rqbox}

\underline{\textbf{Answer for $\mathbf{RQ_1}$-}} AIPRs reshape review work by turning reviewer attention toward a cost-aware intake before investing sustained review effort. Reviewers do not treat AI authorship as unacceptable, but they resist broad, weakly explained, or unvalidated generated changes that shift preparation work to maintainers. Productive AIPR review begins when a responsible human scopes, explains, validates, and owns the change.

\subsection{RQ$_2$- Reviewability Triage for AIPRs}
\label{sec:rq2}

RQ$_2$ investigates how reviewers triage AIPRs before investing detailed effort, including whether passing CI is sufficient, which entry and deferral gates reviewers apply, how they handle suspected drive-by submissions, and which signals make newcomer AIPRs credible.

\subsubsection{Willingness to proceed across change types}

We asked respondents how willing they were to accept CI passed human- and AI-generated revisions without requesting additional explanation. The eight change types covered low-risk routine changes, claim-sensitive changes, and high-risk reasoning-intensive changes. We summarize willingness using the share of respondents selecting \textit{likely} or \textit{very likely}, while inferential tests use the original five-point paired ordinal ratings.

\begin{figure}[htbp]
\centering
\includegraphics[width=14cm]{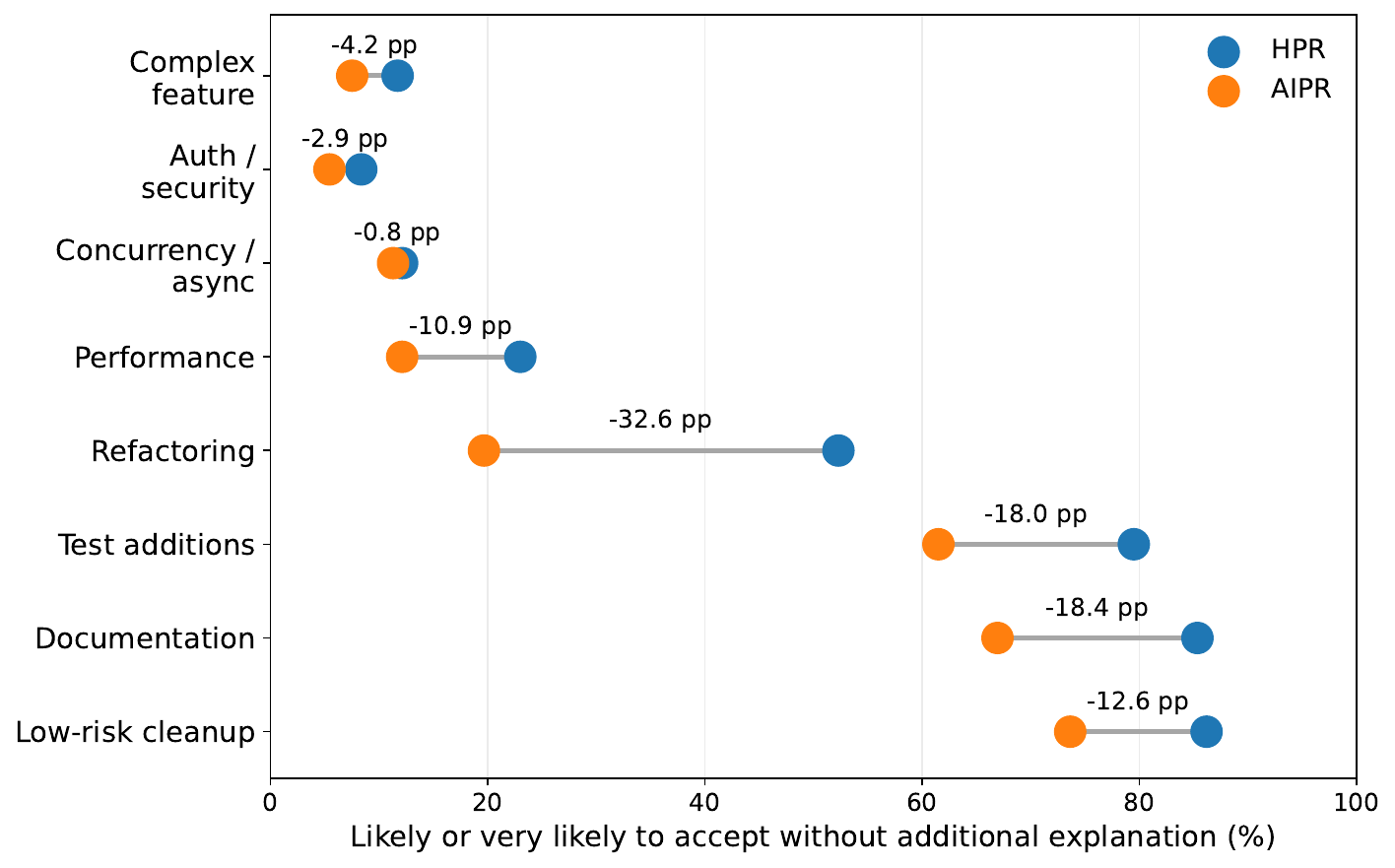}
\caption{Participants reported willingness to accept CI-passing PRs.}
\label{fig:rq2-change-type-willingness}
\end{figure}

Figure~\ref{fig:rq2-change-type-willingness} shows that willingness was significantly lower for AIPRs than HPRs across all eight change types. The gap appeared even for low-risk changes: willingness decreased for cleanup (86.2\% to 73.6\%, $\Delta=-12.6$ pp), documentation updates (85.4\% to 66.9\%, $\Delta=-18.4$ pp), and test additions (79.5\% to 61.5\%, $\Delta=-18.0$ pp), all with $p_{\text{Holm}}<.001$. Thus, passing CI did not make AI-generated revisions as readily acceptable as human-authored revisions, even for routine work. The largest authorship difference appeared for claim-sensitive changes, where correctness depends on preserved behavior, performance effects, or hidden side effects that CI may not expose. For refactors claiming no behavior change, willingness dropped from 52.3\% for HPRs to 19.7\% for AIPRs ($\Delta=-32.6$ pp, $p_{\text{Holm}}<.001$, $r_{rb}=-0.837$), the largest observed drop. Performance changes also decreased from 23.0\% to 12.1\% ($\Delta=-10.9$ pp, $p_{\text{Holm}}<.001$, $r_{rb}=-0.712$). For high-risk changes, willingness was already low in both conditions, but paired ordinal tests still showed significant downward shifts for AIPRs in concurrency or async logic, auth/security-sensitive code, and complex feature logic ($p_{\text{Holm}}<.001$ for all). Therefore, small top-two gaps for high-risk changes should not be read as equivalent acceptance. Reviewers were already reluctant to proceed without explanation, and AI authorship pushed ratings further downward. At the respondent level, more respondents rated the AI-generated revision lower than higher for every change type, from 81 versus 39 for test additions to 146 versus 21 for refactors (Table~\ref{tab:paired}). Because the AIPR item asked about additional \emph{human} explanation and always followed the HPR item, these differences are upper bounds on the effect of authorship alone (Section~\ref{sec:threats}). We also tested whether the AIPR explanation penalty varied by code-review experience and found no evidence that it did. The respondent-level mean penalty showed no significant monotonic association with code-review experience overall ($\rho=0.049$, 95\% CI [$-0.08$, $0.17$], $p=.455$), or within low-, medium-, or high-risk change groups. Kruskal--Wallis tests across the five experience groups were also non-significant after Holm correction. This suggests that the AIPR explanation penalty was broadly distributed rather than driven mainly by less experienced reviewers.

\begin{table*}[htbp]
\centering
\caption{Paired HPR--AIPR comparisons for willingness to accept without additional explanation across change types. Top-two denotes the percentage selecting ``Likely'' or ``Very likely.'' Direction columns report the percentage of respondents whose AIPR rating was lower than, equal to, or higher than their HPR rating ($n=239$).}
\label{tab:paired}
\begin{adjustbox}{width=\textwidth}
\begin{tabular}{p{3.0cm}p{2.8cm}rrrrrrrr}
\toprule
Change type & Risk tier & HPR top-two & AIPR top-two & Gap & $r_{rb}$ & $p_{\text{Holm}}$ & AIPR lower & Same & AIPR higher \\
\midrule
Low-risk cleanup & Low-risk / routine & 86.2 & 73.6 & -12.6 & -0.606 & $<.001$ & 40.2 & 49.8 & 10.0 \\
Documentation & Low-risk / routine & 85.4 & 66.9 & -18.4 & -0.602 & $<.001$ & 50.2 & 36.0 & 13.8 \\
Test additions & Low-risk / routine & 79.5 & 61.5 & -18.0 & -0.419 & $<.001$ & 33.9 & 49.8 & 16.3 \\
Refactor with no behavior change & Claim-sensitive / medium-risk & 52.3 & 19.7 & -32.6 & -0.837 & $<.001$ & 61.1 & 30.1 & 8.8 \\
Performance & Claim-sensitive / medium-risk & 23.0 & 12.1 & -10.9 & -0.712 & $<.001$ & 54.0 & 36.4 & 9.6 \\
Concurrency / async logic & High-risk / reasoning-intensive & 12.1 & 11.3 & -0.8 & -0.739 & $<.001$ & 49.0 & 44.8 & 6.3 \\
Auth / security-sensitive code & High-risk / reasoning-intensive & 8.4 & 5.4 & -2.9 & -0.562 & $<.001$ & 17.2 & 78.2 & 4.6 \\
Complex feature logic & High-risk / reasoning-intensive & 11.7 & 7.5 & -4.2 & -0.723 & $<.001$ & 39.7 & 52.7 & 7.5 \\
\bottomrule
\end{tabular}
\end{adjustbox}
\end{table*}

\begin{rqbox}
\textbf{Finding 2.1: }Even when CI passed, reviewers were less willing to accept AIPRs without additional explanation across all eight change types. The largest drop occurred for behavior-preserving refactors, and the penalty was not reliably moderated by code review experience.
\end{rqbox}

\subsubsection{Entry gates and deferral triggers}

Having shown that CI was not sufficient for AIPR acceptance, we next focus on what makes an AIPR reviewable before reviewers invest effort. We analyze two multi-selection items: stricter entry criteria and deferral or skip triggers. We report selection rates as primary evidence, with 95\% confidence intervals, and use exact McNemar tests to compare options.

\begin{figure*}[htbp]
    \centering
    \includegraphics[width=\linewidth]{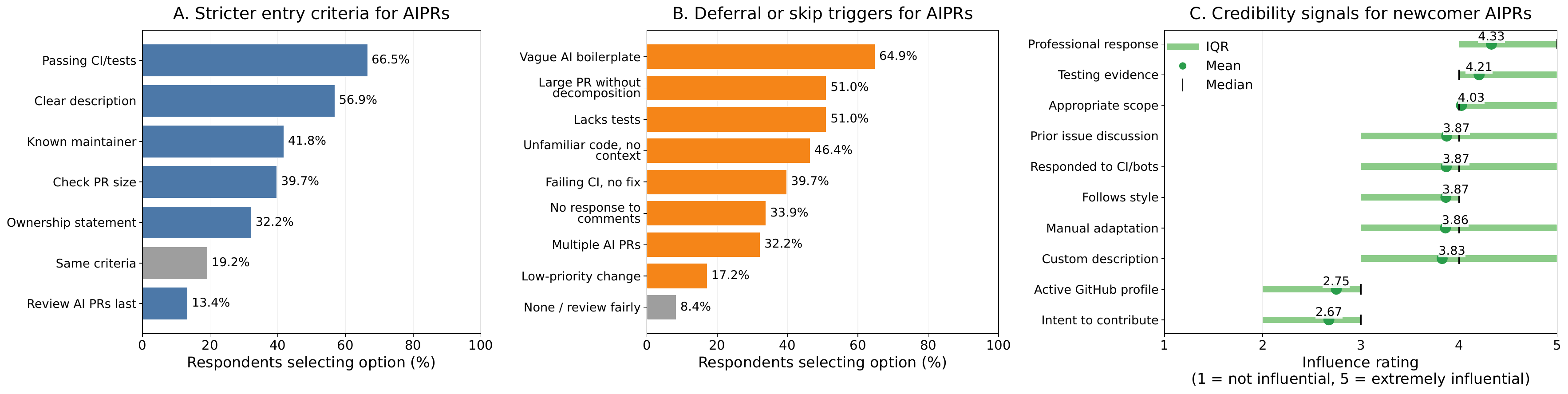}
    \caption{Review-entry gates, deferral triggers, and credibility signals for AIPRs.}
    \label{fig:rq2-aipr-gates-signals}
\end{figure*}

Figure~\ref{fig:rq2-aipr-gates-signals}A shows that most respondents applied at least one additional entry gate before reviewing AIPRs. The most common requirement was passing CI or tests first (66.5\%), followed by a clear PR description or intent (56.9\%). Respondents also looked for accountable ownership: 41.8\% required evidence that a known human would maintain the code, and 32.2\% checked for an explicit ownership statement. Only 19.2\% reported applying the same criteria regardless of authorship, and 13.4\% said they now review AIPRs last. Exact McNemar tests with Holm correction showed that passing CI or tests was selected more often than every other criterion, and a clear description more often than ownership statements, PR size, and a known maintainer, indicating that stricter AIPR triage centered on technical readiness, reviewability, and ownership rather than categorical deprioritization. Deferral triggers followed the same logic (Figure~\ref{fig:rq2-aipr-gates-signals}B). Respondents most often deferred AIPRs with vague or generic AI-generated descriptions (64.9\%), very large PRs without clear decomposition (51.0\%), or missing tests despite behavior changes (51.0\%). Other common triggers were unfamiliar code without context or rationale (46.4\%), failing CI with no attempted fix (39.7\%), and contributors who had not responded to previous review comments (33.9\%). Only 8.4\% selected that they try to review all PRs fairly regardless of these factors. A vague AI-generated description was selected more often than every other trigger (exact McNemar tests, Holm-corrected). These results suggest that reviewers defer AIPRs when submitters shift review cost onto them by omitting explanation, evidence, decomposition, or responsiveness.

Optional open-ended responses provided by 40 respondents for entry gates (16.7\%) and 36 for deferral triggers (15.1\%) reinforced the earlier pattern, and were coded with high agreement ($\kappa=0.97$ and $\kappa=0.98$). Respondents requested risk-specific evidence, such as reproduction notes, failing-before-passing tests, benchmarks, stress traces, screenshots, API-version checks, domain-owner confirmation, threat models, and rollback plans. They also described AIPRs as too costly to review when broad refactors were not split, CI failures were left unexplained, unfamiliar code lacked an owner, or generated submissions were repeatedly opened and abandoned. Overall, reviewers did not reject AIPRs categorically, but required evidence, context, decomposition, and human accountability before investing substantial review effort.

\begin{rqbox}
\textbf{Finding 2.2:} Reviewers use AIPR-specific entry and deferral gates before investing review effort, prioritizing CI/test readiness, clear intent, manageable scope, human ownership, and contributor responsiveness.
\end{rqbox}

\begin{table*}[htbp]
\centering
\caption{Specific team-level rule themes respondents proposed for governing AIPRs. Percentages are over 171 cleaned responses and a response can carry several themes.}
\label{tab:rq3-team-rules}
\resizebox{\linewidth}{!}{%
\begin{tabular}{p{0.20\linewidth}p{0.40\linewidth}p{0.60\linewidth}}
\toprule
\textbf{Rule theme} & \textbf{Problem solved by the rule} & \textbf{Example responses} \\
\midrule
AI-use boundary and admissibility policy (n = 63, 36.8\%) & Clarifies when AI-authored work is acceptable, restricted, or rejected, and distinguishes light AI assistance from agent-generated PRs or low-value generated submissions. & \textit{``[..] guideline should distinguish between AI-assisted editing and fully agent-generated PRs.''}, \textit{``No matter ai or human made, make prs small, it’s manageable then. But prefer only giving boilerplate/tooling to ai.''} \\
\cmidrule{2-3}

Change-type evidence bundles (n = 57, 33.3\%) & Replaces generic requests for tests with evidence tailored to the risk introduced by a generated change, such as explain plans, dry-runs, rollback plans, screenshots, or accessibility checks. & \textit{``I would require generated SQL or migration PRs to include an explain plan, sample table size, and rollback note.''}, \textit{``For AI-authored accessibility changes, I would ask for screen reader or browser accessibility-tree evidence.''} \\
\cmidrule{2-3}

AI-generated volume and queue controls (n = 30, 17.5\%) & Protects reviewer bandwidth from cheap, repeated, or unsolicited AIPR submissions by limiting open PRs, requiring issue-first discussion, and closure or reopen-with-evidence rules. & \textit{``I would set a soft limit on how many AI-authored PRs one contributor can have open at once.''}, \textit{``No unsolicited PRs, ever. It is too easy to add features we don't need.''} \\
\cmidrule{2-3}

Risk-based routing and stakeholder sign-off (n = 26, 15.2\%) & Routes generated changes to the appropriate human authority for security, permissions, migrations, performance, user-facing text, accessibility, legal policy, or cross-subsystem behavior.
 & \textit{``AI-authored changes touching auth, permissions, or secrets require a domain owner review.''}, \textit{``I would require product or legal approval for AI-generated user-facing wording in sensitive areas.''} \\
\cmidrule{2-3}

Generated-diff containment rules (n = 24, 14.0\%) & Keeps generated diffs reviewable by requiring focused behavior changes, staged PRs, coherent commits, narrow refactors, or explicit behavior-preservation invariants. & \textit{``For AI PRs, I would require smaller staged changes when the diff crosses modules.''}, \textit{``I would add a rule that AI-generated refactors need to be narrow and mechanical.''} \\
\cmidrule{2-3}

Agent-workflow hygiene and learning loops (n = 16, 9.4\%) & Makes agentic development governable by documenting prompt or workflow context, aligning agents with project instructions, and feeding review corrections back into future agent behavior. & \textit{``For any comment, which ended up changing the code in a meaningful way, we update the AI's instruction file(s), so this is now documented for the future.''}, \textit{``Use prompt template aligned to company guidelines.''} \\
\bottomrule
\end{tabular}
}
\end{table*}

\subsubsection{Conditional credibility for newcomer AIPRs}

We next characterize suspected AI-assisted drive-by PRs from newcomers, where reviewers face uncertainty about both generated code and contributor commitment. Respondents did not favor automatic closure. Instead, they preferred conditional review: 193 respondents (80.8\%) selected at least one special-handling strategy, while only 40 (16.7\%) selected ``no special handling.'' The most common strategies required the contributor to demonstrate understanding before review (55.6\%), participate in issue discussion before submitting a PR (38.1\%), meet a stricter quality bar than established contributors (33.1\%), or pass automated screening before human review (29.7\%). Each of these strategies was selected significantly more often than no special handling (Holm $p<.001$, paired OR $=1.81$ to $3.35$). By contrast, more punitive or bandwidth-limiting options, such as template closure if imperfect (20.1\%), queue deprioritization (23.8\%), or closing apparently unmodified AI output (12.6\%), did not differ significantly from no special handling after correction. Open-ended responses (37, 15.5\%) reinforced this conditional-review pattern. Respondents described moving PRs to triage until ownership was clear, routing security-sensitive changes to domain owners, tagging a human maintainer, and asking for validation artifacts such as explain plans, migration dry-runs, rollback plans, smaller follow-up PRs, issue-discussion links, or evidence of manual adaptation. Targeted association tests found no robust Holm-significant concentration of these strategies by AIPR-related pressure, code-review experience, or AIPR exposure. Thus, reviewers generally kept a path open for newcomer AIPRs, but expected submitters to establish accountability, evidence, and reviewability before receiving deeper human review.

This conditional-review pattern raises a follow-up question: \textit{what observable signals make a newcomer AIPR credible enough to move from suspected low-effort submission to reviewable contribution?} Respondents rated 10 credibility signals on a five-point ordinal scale. Figure~\ref{fig:rq2-aipr-gates-signals}C shows that the strongest signals were concrete behaviors within the current review process. Professional response to initial feedback was rated highly or extremely influential by 82.4\% of respondents, followed by testing evidence beyond ``it works for me'' (81.2\%), appropriate scope (77.8\%), and response to CI failures or bot comments (71.5\%). By contrast, profile-level or future-commitment cues were much weaker: an active GitHub profile was rated highly or extremely influential by 23.4\%, and stated intent to contribute more or become a maintainer by 21.3\%. A Friedman test confirmed significant differences among signals ($\chi^2(9)=564.14$, $p<.001$, Kendall's $W=.262$), and Holm-corrected post-hoc tests showed that process-level signals were consistently rated higher than profile-level or future-commitment signals. Together, these results show how newcomer AIPRs can become credible without relying on prior reputation. Newcomers reduce reviewer skepticism by making the current PR easy to evaluate: keeping the change focused, explaining intent in project-specific terms, providing testing evidence, responding to CI or bot feedback, adapting generated code to project conventions, and engaging professionally with reviewer comments. AIPRs may lower the barrier to producing a contribution, but they also raise the burden of demonstrating visible human effort before reviewers invest review time.

\begin{rqbox}
\textbf{Finding 2.3:} Reviewers kept a conditional path open for newcomer AIPRs, requiring suspected AI-assisted drive-by submissions to earn credibility through understanding, testing evidence, appropriate scope, CI responsiveness, and professional engagement.
\end{rqbox}

\underline{\textbf{Answer for $\mathbf{RQ_2}$-}} AIPRs become reviewable when they demonstrate readiness beyond passing CI: bounded scope, project-specific rationale, relevant validation, responsive engagement, and human ownership. For newcomers and suspected drive-by submissions, credibility depends on visible participation in the current review, not profile reputation or stated future commitment. Overall, AIPR triage protects reviewer effort while preserving a path for accountable AI-assisted contributions.

\subsection{RQ$_3$- Fairness and Governance in AIPR Workflows}
\label{sec:rq3}

RQ$_3$ focuses on how reviewers preserve fairness and accountability in AIPR workflows. We organize the analysis around two connected concerns: the project-level rules reviewers want for AIPR intake, and the social conditions under which stricter AIPR handling remains fair.

\begin{table*}[htbp]
\centering
\caption{Themes in open-ended responses about effort asymmetry in AIPR review. Percentages are over 128 codeable responses and a response can carry several themes.}
\label{tab:rq3-review-reciprocity}
\resizebox{\linewidth}{!}{%
\begin{tabular}{p{0.20\linewidth}p{0.40\linewidth}p{0.60\linewidth}}
\toprule
\textbf{Theme} & \textbf{What it explains} & \textbf{Example responses} \\
\midrule
Accountable human stewardship($n=78$, 60.9\%) & AI use becomes acceptable when the human submitter visibly understands, can validate and explain change rationale, and responds as an accountable participant. & \textit{``If the contributor clearly validates and explains the AI-generated change, I treat it much more like a normal contribution.''}, \textit{``If they show that care, I will show similar care.''} \\
\cmidrule{2-3}

Reciprocal reviewer investment ($n=73$, 57.0\%) & Reviewers are more willing to invest time, explain reasoning, and mentor when they see that the contributor invested effort, while low-effort submissions weaken this sense of reciprocity. & \textit{``When I know someone invested significant human effort in writing the code, I usually feel more willing to spend time [.....] and giving detailed feedback.''}, \textit{``When someone worked hard on something I feel more like I should put in the effort to give real feedback.''} \\
\cmidrule{2-3}

Audit mode for prompt-only AIPRs ($n=58$, 45.3\%) & AIPRs with little visible human work make review feel more transactional, risk-focused, and less mentoring-oriented. & \textit{``With AI-generated stuff I'm more just checking whether it's safe to merge.''}, \textit{``I experience it as a document to audit rather than a collaboration to participate in.''} \\
\cmidrule{2-3}

Offloaded strain and frustration ($n=55$, 43.0\%) & Reviewers feel that low-effort AI submissions can shift debugging, edge-case reasoning, and validation work onto them. & \textit{``It feels like they are treating the reviewer as a human debugger to clean up after their bot.''}, \textit{``If it saves the author ten minutes and costs the reviewer an hour, it feels unfair and unsustainable.''} \\
\cmidrule{2-3}

Tool-neutral review standards ($n=22$, 17.2\%) & A smaller group emphasized that review should focus on correctness, quality, and ownership rather than whether the code was manually written or AI-assisted. & \textit{``I try to evaluate code based on quality and correctness rather than how it was produced.''}, \textit{``If they have put the effort in to understand the change and worked to ensure its quality, it does not matter who typed it.''} \\
\bottomrule
\end{tabular}
}
\end{table*}

\subsubsection{Governance rules for AIPR intake}

We asked respondents what one team guideline they would change to better handle AIPRs and what problem the change would solve. The goal was to identify concrete workflow rules beyond individual reviewer judgment. We analyzed 171 cleaned responses using multi-label thematic coding because many responses proposed more than one rule. During open coding ($\kappa = 0.91$), we identified 24 rule subthemes and grouped related subthemes into six higher-level rule families. We coded this item at the level of concrete workflow mechanisms rather than broad attitudes because RQ$_1$ and RQ$_2$ already captured general reviewability, evidence, and accountability concerns.

Table~\ref{tab:rq3-team-rules} summarizes the resulting rule themes, showing that respondents wanted AIPR governance embedded in project workflow infrastructure rather than left to reviewer caution alone. They proposed concrete rules for AI-use boundaries and admissibility policies (63 responses, 36.8\%), clarifying when AI use must be disclosed, accepted, or restricted. The second most prevalent theme was change-type evidence bundles (57, 33.3\%), where respondents asked teams to replace generic test requirements with evidence matched to the generated change, such as explain plans, migration dry-run notes, UI screenshots, or performance benchmarks. Other themes addressed the workflow burden of generated submissions, including queue controls (30, 17.5\%), risk-based routing to domain owners or stakeholders (26, 15.2\%), generated-diff containment rules (24, 14.0\%), and agent-workflow hygiene (16, 9.4\%). Overall, respondents wanted AIPR governance encoded into PR templates, review queues, owner-routing rules, agent instructions, and admissibility policies.

\begin{rqbox}
\textbf{Finding 3.1:} Reviewers proposed workflow-specific governance rules for AIPRs, especially AI-use admissibility policies and risk-specific evidence requirements, followed by queue controls, owner routing, generated-diff containment, and agent-workflow hygiene.
\end{rqbox}

\begin{table*}[htbp]
\centering
\caption{Fairness triggers in open-ended responses and how they transfer to AIPRs. Percentages are over 129 codeable responses and a response can carry several triggers.}
\label{tab:rq3-fairness-triggers}
\resizebox{\linewidth}{!}{%
\begin{tabular}{p{0.25\linewidth}p{0.50\linewidth}p{0.60\linewidth}}
\toprule
\textbf{Fairness trigger} & \textbf{How it works in review} & \textbf{How it transfers to AIPRs} \\
\midrule
Visible effort, newcomer status, or learning intent ($n=76$, 58.9\%) & Reviewers become more patient and mentoring-oriented when a newcomer, junior developer, or learner clearly invested effort. & The trigger transfers only when the human behind the AIPR shows learning intent and can explain the generated change in their own words. \\
\cmidrule{2-3}

Engaged communication and humility ($n=63$, 48.8\%) & Thoughtful descriptions, uncertainty statements, follow-up questions, responsiveness, and willingness to revise make reviewers feel obligated to give constructive feedback. & For AI generated changes, reviewers look for the submitter's behavior between review rounds, not just the polish of the generated code. Mechanical fixes without understanding weaken the trigger. \\
\cmidrule{2-3}

Validation evidence and reproducible context ($n=57$, 44.2\%) & Tests, screenshots, explain plans, rollback notes, issue links, commands run, and other validation artifacts show respect for reviewer time. & The same evidence matters more for AIPRs because reviewers cannot infer human understanding from generated code. Validation must be human-owned rather than merely generated. \\
\cmidrule{2-3}

Risk, ownership, and stakeholder context ($n=51$, 39.5\%) & Reviewers become more careful when changes affect security, billing, deployment, accessibility, customer incidents, or user-facing policy. & For AIPRs, these triggers require an accountable human owner or stakeholder because the agent cannot own production, policy, or post-merge consequences. \\
\cmidrule{2-3}

Prior relationship and community stewardship ($n=24$, 18.6\%) & Familiar contributors, trusted teammates, and community members who are likely to remain engaged receive more patient and constructive review. & The trigger is weaker for anonymous or drive-by AIPRs unless the submitter demonstrates continued engagement and responsibility. \\
\cmidrule{2-3}

Tool-neutral fairness standard ($n=20$, 15.5\%) & Some respondents argued that all PRs should receive the same quality bar regardless of tool use, contributor status, or authorship. & Even in this group, several respondents still expected the human submitter to understand and own the AI-generated result before review care could be extended. \\
\bottomrule
\end{tabular}
}
\end{table*}

\subsubsection{Fairness under effort asymmetry}

The intake rules above show (Table \ref{tab:rq3-team-rules}) what reviewers want projects to require. We next analyze why those rules matter socially. Respondents described code review as a reciprocal exchange of effort: reviewers invest more constructively when submitters visibly understand, validate, and own their changes, but shift toward audit, triage, or rejection when AIPRs appear to offload explanation, debugging, or validation onto maintainers. We analyzed two open-ended items to characterize this fairness logic. First, 137 respondents discussed how reviewing substantial human effort versus prompt-produced code affected their willingness to invest review time, emotional response, and sense of reciprocity. After excluding brief or low-information entries, 128 responses were codeable. Open coding produced 14 subthemes grouped into five mechanisms ($\kappa=0.81$). Second, 136 respondents described when they feel a stronger obligation to treat a submitter fairly or carefully, and whether the same trigger applies to HPRs and AIPRs. Of these, 129 responses were codeable. Open coding produced 18 trigger subthemes grouped into six trigger families ($\kappa=0.84$). Tables~\ref{tab:rq3-review-reciprocity} and~\ref{tab:rq3-fairness-triggers} show a consistent pattern. Visible human effort increased patience, detailed feedback, and mentoring investment, while prompt-only AIPRs often moved reviewers into audit mode. Reviewers did not reject AI use uniformly. They described the asymmetry as repaired when the submitter explained intent, validated the output, responded to feedback, and accepted post-merge responsibility. Similarly, fairness triggers such as newcomer status, uncertainty, preparation, and learning transferred to AIPRs only when human engagement was visible. Because generated code can look polished regardless of the submitter's understanding, reviewers looked for human-side signals around the PR: explanation, uncertainty disclosure, validation, responsiveness, and ownership. This pattern reconciles the tension between fairness and stricter AIPR handling. Reviewers did not define fairness as identical treatment of every submission. Instead, they treated fairness as a conditional obligation shaped by evidence of human engagement. When submitters demonstrated learning, ownership, and responsiveness, reviewers were more willing to be patient and constructive. When an AIPR appeared to be low-effort prompt output that transferred work to maintainers, reviewers narrowed their response to audit, triage, or rejection.

\begin{rqbox}
\textbf{Finding 3.2:} Stricter AIPR handling was viewed as fair when it addressed effort asymmetry, not AI authorship alone. Reviewers extended mentoring when human engagement was visible, but shifted to audit, triage, or rejection when AIPRs offloaded work onto maintainers.
\end{rqbox}

\underline{\textbf{Answer for $\mathbf{RQ_3}$-}} Reviewers expect AIPR governance to make human stewardship visible before ordinary review. They want rules for acceptable AI use, risk-specific evidence, generated-diff containment, owner routing, and queue control. Stricter handling is viewed as fair when it protects reviewer effort while keeping a clear path for accountable contributors.

\section{Discussion}
\label{sec:discussion}

\subsection{A Conditional Review Model for AIPRs} 

Reviewers are not deciding whether AI-authored work should be rejected as a category, but whether a generated contribution exposes enough human accountability to justify scarce review attention. This extends prior work showing that PR evaluation depends on technical signals, contributor history, project membership, and interaction quality~\cite{gousios2014exploratory,tsay2014influence,bacchelli2013expectations,kononenko2016code}. AIPRs add a new condition to this social contract: artifact quality no longer reliably implies human understanding. A polished diff, fluent description, or passing CI result may still leave unclear who shaped the change, checked its assumptions, and will maintain it after merge. Thus, reviewing AIPRs requires reviewers to calibrate their effort alongside their trust in AI-generated contributions. Human-AI trust research argues that reliance should be calibrated to task risk, uncertainty, and available evidence~\cite{parasuraman1997humans,lee2004trust,amershi2019guidelines}. AIPR review requires parallel calibration of reviewer effort: how much reasoning reviewers must supply before generated code becomes acceptable. This explains why reviewers can be open to AIPRs while still requiring stronger intake signals before sustained review begins. 

\subsection{Cheap Signals, Costly Signals, and Reciprocity} 

The lenses in Section~\ref{sec:lens} help explain this pattern. In signaling terms, generative AI has made several traditional signals of contributor effort cheap. A fluent description no longer shows that the contributor understood the change, and a vague or generic AI-generated description was the most frequent deferral trigger (64.9\%). Respondents also gave little weight to signals that say nothing about the current change: an active GitHub profile and a stated intent to keep contributing were rated highly or extremely influential by only 23.4\% and 21.3\%, although activity traces and profiles have been an important basis for impressions on GitHub~\cite{marlow2013impression}. The signals they valued most concern the contributor's behavior in the current review, such as professional responses to initial feedback (82.4\%) and testing evidence (81.2\%). Responding well to feedback, in particular, is hard to do without understanding the change. Similarly, the most common way of handling suspected drive-by PRs was to ask the contributor to demonstrate understanding before review (55.6\%). In social exchange terms, 57.0\% of the codeable effort-asymmetry responses described reciprocal reviewer investment: reviewers offered explanation and mentoring in proportion to the effort they saw. Finally, respondents judged stricter handling fair when it responded to missing evidence or effort rather than to AI authorship (Finding~3.2). In justice terms, this favors procedural fairness, because rules about observable behavior can be applied consistently to every contributor~\cite{colquitt2001dimensionality,german2018my}. However, these interpretations are post hoc and should be examined in future studies using designs that manipulate the relevant signals directly, such as vignette experiments.

\subsection{Reviewability Debt and Recoverable Fairness} 

We term the remaining gap \emph{reviewability debt}: the cost transferred to reviewers when a PR appears complete but leaves intent, scope, validation, or ownership unresolved~\cite{bosu2015characteristics,yu2015wait}. Unlike technical debt~\cite{tom2013exploration,murillo2023identification}, its immediate cost is paid at the review interface through defensive checking, repeated clarification, queue delay, or reduced mentoring. The concept also explains why CI is necessary but insufficient~\cite{mcintosh2016empirical,spadini2018testing}: CI can show that checks passed, but not whether a generated refactor preserved behavior, a performance claim is meaningful, or a security-sensitive change was understood by an accountable human. Reviewability debt also changes fairness. Prior work shows that fairness shapes contributor experience, while newcomer success depends on learning project norms, receiving guidance, and responding constructively~\cite{german2018my,steinmacher2015social,tsay2014influence}. AIPRs complicate these norms because generated code can conceal whether the submitter is learning or merely forwarding model output. Fair AIPR review should therefore mean \emph{recoverable access}, not identical treatment: contributors receive a clear path into review, while maintainers are not expected to absorb unlimited validation and explanation work. Issue-first discussion, template closure, owner requests, and reopen-with-evidence rules can be fair when they explain what is missing and how to proceed. 

\subsection{Designing for Stewardship at PR Intake} 

Disclosure alone is too weak as an AIPR governance mechanism. Projects need a \emph{stewardship trace}: a visible chain from generated output to human judgment that shows what was generated, what was manually revised, what validation was run, what assumptions remain open, and who will maintain the change. This connects to work cautioning that repository events may not reveal meaningful human oversight, and to developer-centered AI studies calling for bounded delegation, provenance, authority scoping, and uncertainty signaling~\cite{duma2026these,choudhuri2026copilot}. For AIPRs, these principles should be embedded before ordinary review begins. For tool builders, the design target should be reviewability, not only code generation. Coding agents should produce reviewability artifacts alongside code, including provenance summaries, risk-specific evidence, validation logs, and ownership declarations. This aligns with human-AI design guidance on visibility, user control, and appropriate reliance, and with evidence that repository traces alone may not capture human oversight in AI-generated PRs~\cite{amershi2019guidelines,choudhuri2026copilot,duma2026these}. PR platforms should support generated-diff containment, human-edited rationale, and routing to responsible owners for sensitive changes. For practitioners, policies should avoid one-size-fits-all AI rules: routine generated changes may need lightweight checks, while refactors, migrations, security, performance, accessibility, and user-facing changes require stronger evidence and ownership. This intake layer should also preserve newcomer access. Since review fairness shapes contributor experience and newcomers often need explicit guidance to learn project norms~\cite{german2018my,steinmacher2015social,tsay2014influence}, clear entry rules, issue-first discussion, and reopen-with-evidence mechanisms can teach responsible AI-assisted contribution while giving maintainers a principled way to defer poorly prepared work. The goal is not to make AIPRs harder to submit, but to make responsibility visible before scarce reviewer effort is spent. 

\subsection{From Project Policies to Reviewer Judgments}

Our results converge with recent studies that used other kinds of evidence. The rationales that \citet{robles2026you} found in project AI policies, protecting maintainer attention, preserving accountability, sustaining meaningful review interaction, and keeping pathways open for newcomers, correspond to our respondents' deferral triggers, ownership gates, reciprocity theme, and preference for conditional handling of newcomer AIPRs. The review friction and workload asymmetry that \citet{baltes2026endless} found in public discussions match the effort asymmetry our respondents described (Section~\ref{sec:rq3}). Two differences stand out. First, policy studies report mainly whether AI may be used, whether its use must be disclosed, and whether a human must be involved and accountable~\cite{hora2026ai,hora2026we}. Our respondents also asked for rules on what an AIPR must contain and how it is handled: evidence matched to the type of change (33.3\%), routing risky changes to domain owners or other stakeholders (15.2\%), and keeping generated diffs focused (14.0\%). Second, aggressive PR closure was the most common countermeasure in project policies~\cite{hora2026we}, and \citet{afroz2026ai} found that communities often default to defensive strategies. In contrast, when asked about suspected AI-assisted drive-by PRs from newcomers, only 12.6\% of our respondents would close PRs that appear to be unmodified AI output, and the most common strategies asked the contributor to demonstrate understanding (55.6\%) or to discuss the issue first (38.1\%). Individual reviewers may prefer to keep a path open, whereas projects facing high review volume may formalize closure practices. The gap may also partly reflect social desirability in self-reported responses (Section~\ref{sec:threats}).

\subsection{Research Agenda}

Our findings suggest three directions for future work. First, longitudinal studies should examine the post-merge life of AIPRs by linking provenance, review-time evidence, and later outcomes such as defects, refactoring effort, and other maintenance. Second, controlled and field studies should investigate how AIPRs affect reviewer cognition, including automation bias, fatigue, calibration, and skill development, especially for junior and mid-level reviewers. Third, future work should evaluate which evidence actually improves AIPR review, including provenance traces, validation logs, rationale records, human-adaptation signals, and AI-to-AI review. The central research challenge is to understand when generated contributions provide independent, verifiable assurance rather than shifting uncertainty to human reviewers.

\section{Threats to Validity}
\label{sec:threats}

\subsection{Construct Validity}

The constructs examined in this study, including reviewability, fairness, and accountability, were operationalized through study-specific questionnaire items
rather than established psychometric scales. As described in
Section~\ref{sec:survey-design}, the instrument was grounded in prior research and refined through expert review and pilot testing. We therefore interpret the individual items as indicators of reviewers' stated judgments rather than as
validated latent measures of these constructs. Confidence in the resulting interpretations comes from convergence across structured responses and open-ended explanations.

An important measurement choice concerns what constitutes an AIPR. The study captures \emph{perceived authorship} rather than verified line-level provenance. In contemporary development workflows, AI generation, human prompting, manual editing, validation, and human submission can be intertwined, making a strict boundary between AI-assisted and AI-authored code difficult to observe during review. The questionnaire nevertheless provided a consistent context by referring to PRs substantially produced using AI-based code generation systems and by distinguishing the generated contribution from the human responsible for prompting, validating, submitting, or maintaining it. The open-ended responses also consistently discuss generated code, prompting,
AI-use disclosure, human validation, and post-merge ownership. We therefore interpret AIPR throughout the paper as a PR that respondents perceived to be substantially produced by AI, while acknowledging that individual respondents may have applied somewhat different thresholds for that perception. AIPR exposure was likewise self-reported in ordinal bands and is therefore more appropriate for characterizing the sample and exploratory comparisons than for precise exposure measurement.

Predefined response alternatives can make listed considerations more salient
than alternatives respondents must generate independently
\cite{krosnick1991response,krosnick1999survey}. As described in
Section~\ref{sec:survey-design}, the questionnaire included opportunities to
report considerations beyond the predefined options. Importantly, open-ended
questions about review-practice changes and desired team guidelines preceded
the more detailed lists of entry criteria, deferral conditions, and newcomer
signals. These unconstrained responses independently surfaced concerns involving
validation, intent, ownership, scope, and reviewer bandwidth that also appeared
in the structured items. This convergence supports the relevance of the
predefined alternatives, but not their exhaustiveness. We therefore interpret
selection frequencies as endorsement of the alternatives presented rather than
as estimates of what reviewers would identify spontaneously.

Some item wording intentionally made particular review situations salient. For example, the questionnaire referred to low-effort AI-generated contributions, stricter review criteria, suspected drive-by submissions, and differences in visible human effort. Such formulations allow us to examine how reviewers reason when those situations are present, but they may also influence the framing through which respondents consider AI-assisted contribution. The questionnaire provided counter-positions, including no observed problem, authorship-independent review criteria, and no special handling, and these positions were selected and also appeared in the qualitative responses. We therefore interpret these results as \emph{conditional judgments under the stated scenarios}, rather than as estimates of how often the same concerns would emerge in an unprompted setting. Finally, the study captures respondents' reported experiences and judgments
rather than behavior reconstructed from repository histories. Self-reports can
differ from actual practice and can be affected by recall and social
desirability~\cite{devanbu2016belief,tourangeau2000psychology}. We mitigate
this concern by asking about concrete review situations and practices,
retaining countervailing and tool-neutral responses, and using open-ended
explanations to contextualize structured answers. The findings should therefore
be interpreted as evidence about reviewers' perceptions and stated judgments,
rather than direct observations of review behavior. Similarly,
\emph{reviewability debt} is an interpretive concept synthesized from recurring
patterns in the responses rather than a directly measured quantity of reviewer
time, downstream defects, or maintenance cost.

\subsection{Internal Validity}
\label{sec:threats-internal}

The study is cross-sectional and observational and was designed to characterize reviewer judgments rather than establish a causal effect of AI authorship on review behavior. The situations respondents recalled may differ in generated code quality, change risk, repository maturity, subsystem familiarity, contributor history, organizational AI policy, and prior experience with AI tools. We therefore avoid interpreting the reported differences as effects caused solely by AI authorship.

The paired HPR--AIPR questions strengthen comparison by asking the same respondent to evaluate both conditions. In the willingness block, the same change types and the same assumption that CI passes were used for both HPRs and AIPRs, reducing between-person variation and holding important aspects of the scenario constant. However, the HPR condition preceded the corresponding AIPR condition and the order was not randomized. The comparison may therefore be affected by order, contrast, or increased authorship salience \cite{schuman1996questions}. The wording also placed somewhat greater emphasis on the need for human explanation in the AIPR condition. We therefore interpret the paired results as within-respondent contrasts under an explicit authorship comparison rather than as causal estimates of authorship in natural review settings.

The broader questionnaire context may also have influenced later responses.
Because the questionnaire followed a fixed order, earlier items may have made
certain considerations more salient when respondents answered subsequent
questions. However, the open-ended responses introduced a broader range of
concerns than those explicitly prompted earlier in the questionnaire. We
therefore cannot rule out context or order effects, although the observed
patterns were not limited to considerations made salient by preceding items.

The study description also made the comparison between human- and AI-authored review explicit. Such transparency was necessary for informed participation, but it may create demand characteristics if respondents infer the kinds of differences the researchers expect~\cite{orne2017social}. The response patterns were nevertheless not uniformly unfavorable toward AIPRs: substantial groups reported equivalent treatment, selected authorship-neutral criteria, and described conditions under which AIPRs should receive ordinary review. The
observed differences also varied across change types and review situations. These patterns provide some assurance against a simple uniform demand effect, but they cannot eliminate it. Our conclusions therefore emphasize the conditional structure of reviewers' judgments rather than treating every HPR--AIPR difference as an isolated effect of authorship.

\subsection{External Validity}
\label{sec:threats-external}

We used complementary recruitment sources to obtain practitioners with varied review backgrounds and opportunities for exposure to AI-assisted development. Participants were recruited from repositories with known agent-authored PR activity, from mature Apache and GitHub projects selected independently of known AIPR activity, and through LinkedIn to broaden participation beyond repository-derived contacts. The final sample contains 239 respondents from 31 countries and spans a wide range of software-development and code-review experience. Most respondents also reported substantial prior exposure to AIPRs, providing an experience base relevant to the phenomenon under study.

The recruitment strategy is nevertheless purposive and non-probabilistic. Repository-based recruitment favors practitioners visible in mature open-source projects and those with publicly available contact information, whereas private repositories, smaller projects, regulated organizations, non-GitHub ecosystems, and organizations with different review processes may
be less represented. Professional-network dissemination broadens the sampling frame but also introduces self-selection. Geographic participation is similarly broad but uneven. Following recommendations for sampling in software engineering, we therefore use the sample to support analytical generalization to comparable review settings rather than statistical estimates for the entire software-development population~\cite{baltes2022sampling,ralph2020empirical}. Participants varied in the extent of their experience reviewing AIPRs, but the
achieved sample had substantial direct exposure. In particular, 78.2\% reported
reviewing at least 10 AIPRs and 49.4\% reported reviewing more than 40.
Respondents with lower exposure may reflect emerging expectations rather than
well-established review routines. We therefore report AIPR exposure explicitly
and treat exposure-based analyses as contextual rather than as a basis for
population-level generalization.

Self-selection and nonresponse provide an additional boundary. Recruitment combined direct email invitations with open dissemination through LinkedIn, so the number of individuals who encountered the survey cannot be determined reliably and anonymous responses cannot be assigned to recruitment channels. A single invitation-level response rate would therefore be misleading.
Practitioners who chose to participate may differ from non-participants in AI exposure, workload, organizational policy, or strength of opinion. We address this through transparent reporting of the recruitment process and achieved sample rather than weighting against an unknown population distribution.

The findings are also temporally situated. Coding agents, provenance mechanisms, contribution policies, and reviewer familiarity with generated code are evolving rapidly. Exact selection rates may therefore change as tools and governance practices mature. Replication across industrial settings, repository ecosystems, organizational policies, and later generations of
coding agents will be needed to establish the longer-term stability and transferability of the reported patterns.

\subsection{Conclusion Validity}
\label{sec:threats-conclusion}

The quantitative analyses were matched to the measurement scale and dependency structure of the questionnaire. Within-respondent ordinal comparisons were analyzed using paired non-parametric procedures and accompanied by effect-size measures. Repeated ordinal ratings were analyzed using methods appropriate for
within-respondent comparisons, and Holm correction was applied within families of related tests. We additionally report response distributions and direction-of-change information so that interpretation does not depend on statistical significance alone.

Some analyses, particularly those involving AIPR exposure, software-development experience, and code-review experience, are exploratory. Smaller subgroups and sparse categories reduce statistical precision, so non-significant results are not interpreted as evidence that no association exists. Similarly, selection frequencies from multiple-selection questions are
not treated as a complete ranking of reviewer concerns, especially where respondents were limited to a fixed number of choices.

We also considered response completeness and data quality. From 261 received records, one non-consenting response and 21 responses with more than 50\% of the questionnaire unanswered were excluded, resulting in 239 responses for analysis. Optional and open-ended questions were analyzed using their item-specific denominators, and missing responses were not interpreted as
absence of the corresponding view. Because the survey was anonymous and did not include dedicated attention-check or automated-response detection items, undetected inattentive, repeated, or machine-assisted responses cannot be ruled out completely. However, the questionnaire required substantive engagement with several open-ended questions, participation was not incentivized, and the open-ended responses were subsequently examined during qualitative coding. Responses that were too brief, unclear, or otherwise not interpretable were not assigned substantive themes.

For the qualitative analysis, two coders independently applied a shared
codebook with multi-label coding permitted. Disagreements were resolved through
discussion, with unresolved cases adjudicated by a third evaluator. Cohen's
$\kappa$ indicated strong agreement across the independently coded items,
supporting the consistency of category assignment while recognizing that
qualitative coding remains interpretive. Open-ended survey responses also offer
less opportunity for probing and clarification than interviews. We therefore
use the qualitative findings to explain, contextualize, and qualify the
quantitative patterns, including identifying counterexamples, rather than to
claim theoretical saturation or estimate population prevalence.

\section{Conclusion}
\label{sec:conclusion}

As AI-generated contributions become easier to produce, code review increasingly depends on whether the surrounding human judgment is visible enough for reviewers to act on. Our study suggests that the central governance challenge is therefore not AI authorship itself, but whether generated contributions arrive with sufficient evidence of understanding, validation, scope control, responsiveness, and responsibility. This perspective connects technical reviewability with the social conditions under which reviewers decide to invest attention, provide mentoring, or defer a contribution. We capture the resulting burden through the concept of \emph{reviewability debt}: work left to reviewers when rationale, validation, context, or ownership remain unresolved. This shifts the design problem from detecting AI-generated code toward making stewardship visible at review intake. Tools and project workflows should therefore support evidence that helps reviewers assess whether a generated contribution is ready for sustained human attention, while still preserving a clear path for responsible AI-assisted contributors.

Future work should test these mechanisms beyond self-reported judgments. Repository and longitudinal studies can examine whether reviewability signals predict review latency, rework, maintenance effort, or post-merge outcomes, while controlled and field studies can evaluate which forms of provenance, validation evidence, rationale, and ownership actually reduce reviewer burden. As coding agents improve, the sustainability of AI-assisted development will
depend increasingly on how well software workflows connect generated code to observable human responsibility.

\section*{Declarations}
\subsection*{Funding}

This research is supported in part by the Natural Sciences and Engineering Research Council of Canada (NSERC) Discovery Grants Program, the Canada Foundation for Innovation’s John R. Evans Leaders Fund (CFI-JELF), and the industry-stream NSERC CREATE program in Software Analytics Research (SOAR).

\subsection*{Ethical approval}
This study received ethics approval from the University of Saskatchewan Behavioural Research Ethics Board (Approval No. BEH\#6412) prior to data collection.

\subsection*{Informed consent}
Informed consent was obtained from all participants before participation in the
study.

\subsection*{Author Contributions}

Md Shamimur Rahman led the conceptualization and design of the study, developed the survey instrument, conducted participant recruitment and data collection, performed the quantitative and qualitative analyses, interpreted the findings, and drafted the manuscript. Khairul Alam contributed to the qualitative coding and validation of the open-ended responses and reviewed the manuscript. Banani Roy provided research guidance, contributed to the interpretation of the findings, and critically reviewed the manuscript. Chanchal K. Roy supervised the overall research, provided methodological and strategic guidance, contributed to the interpretation of the findings, and critically reviewed and revised the manuscript. All authors reviewed and approved the final version of the manuscript.

\subsection*{Data Availability Statement} Our replication package and supplementary materials can be found in the following repository: \url{https://doi.org/10.5281/zenodo.21053770}

\subsection*{Conflict of Interest}
We have no conflict of interest.

\subsection*{Clinical Trial Number on the title page}
``Not applicable"

\bibliography{no_dup}

\end{document}